\documentclass{emulateapj}

\newcommand{\GHz}{$\>$GHz }
\newcommand{\MHz}{$\>$MHz }
\newcommand{\Jy}{$\>$Jy}

\shorttitle{The Luminous Hotspot of PKS~1421--490}
\shortauthors{Godfrey et al.}

\begin{document}

\title{A Multi-Wavelength Study of the High Surface Brightness Hotspot in PKS~1421--490}

\author{L.E.H. Godfrey\altaffilmark{1,2}, G.V. Bicknell\altaffilmark{1},
  J.E.J. Lovell\altaffilmark{2,3,4}, D.L. Jauncey\altaffilmark{2},
  J. Gelbord\altaffilmark{5}, \\ D. A. Schwartz\altaffilmark{6},
  H. L. Marshall\altaffilmark{7}, M.~Birkinshaw\altaffilmark{6,8},
  M. Georganopoulos\altaffilmark{9,10}, 
D.W.Murphy\altaffilmark{11}, \\
E.S.~Perlman\altaffilmark{9},
D. M. Worrall\altaffilmark{6,8}}

\email{lgodfrey@mso.anu.edu.au}

\altaffiltext{1}{Research School of Astronomy and Astrophysics, Australian National
University, Cotter Road, Weston, ACT, 2611, Australia}
\altaffiltext{2}{Australia Telescope National Facility, CSIRO, P.O. Box 76, Epping, NSW,
2121, Australia}
\altaffiltext{3}{CSIRO, Industrial Physics, PO Box 218 Lindfield NSW 2070, Australia}
\altaffiltext{4}{School of Mathematics and Physics, University of Tasmania, Tas 7001, Australia}
\altaffiltext{5}{Department of Physics, Durham University, South Road, Durham, DH1 3LE, UK}
\altaffiltext{6}{Harvard-Smithsonian Center for Astrophysics, 60 Garden Street, Cambridge, MA 02138, USA}
\altaffiltext{7}{Kavli Institute for Astrophysics and Space Research, Massachusetts Institute of Technology, USA}
\altaffiltext{8}{Department of Physics, University of Bristol, Tyndall Avenue, Bristol BS8 1TL, UK}
\altaffiltext{9}{Department of Physics, Joint Center for Astrophysics, University of Maryland-Baltimore County, 1000 Hilltop Circle, Baltimore, MD 21250, USA}
\altaffiltext{10}{NASA Goddard Space Flight Center, Mail Code 660, Greenbelt, MD 20771, USA}
\altaffiltext{11}{Jet Propulsion Laboratory, 4800 Oak Grove Drive, Pasadena, CA 91109, USA}

\begin{abstract}

Long Baseline Array imaging of the z=0.663 broad
line radio galaxy PKS~1421--490 reveals a 400 pc diameter high
surface brightness hotspot at a projected distance of approximately 40~kpc
from the active galactic nucleus. The isotropic X-ray luminosity of the hotspot, $L_{2-10\> \rm{keV}} =
3 \times 10^{44} \> \rm{ergs \> s^{-1}}$, is comparable to the isotropic X-ray
luminosity of the entire X-ray jet of PKS~0637--752, and the peak radio surface brightness is hundreds of times greater than that of the brightest hotspot in Cygnus A. We model the radio to X-ray spectral energy distribution using a  one-zone synchrotron self Compton model
with a near equipartition magnetic field strength of 3 mG. There is a strong brightness asymmetry between the approaching and receding hotspots and
the hot spot spectrum remains flat ($\alpha \approx 0.5$) well beyond the predicted
cooling break for a 3~mG magnetic field,
indicating that the hotspot emission may be Doppler beamed. A high plasma velocity beyond the terminal jet shock could be the result of a dynamically important magnetic field in the jet. There is a change in the slope of the hotspot radio spectrum at GHz frequencies from $\alpha \sim 0.5$ to $\alpha \lesssim 0.2$, which we model by incorporating a cut-off in the electron energy distribution  at $\gamma_{\rm{min}} \approx 650$, with higher values implied if the hotspot emission is Doppler beamed. We show that a sharp decrease in the electron number density below a Lorentz factor of 650 would arise from the dissipation of bulk kinetic energy in an electron/proton jet with a Lorentz factor $\Gamma_{\rm jet} \gtrsim 5$.

\end{abstract}


\section{Introduction} \label{sec:intro}

PKS~1421--490 was first reported as a bright, flat spectrum radio source by
\citet{ekers69}. Subsequent VLBI imaging revealed 10mas scale 
structure within the brightest component of this source
\citep{preston89}. Studies at the Australia Telescope Compact Array (ATCA)
later revealed significant radio emission on arcsecond scales extending
south-west from the brightest component \citep{lovell97}. For this reason,
PKS~1421--490 was  included in a {\it Chandra} survey of flat spectrum radio
quasars with arcsecond scale radio jets \citep{marshall05}. \citet{gelbord05}
(from here on G05) reported on recent X-ray ({\it Chandra}), optical (Magellan) and
radio (ATCA) imaging of this source. We refer the reader to that paper for
the details of these observations and images.  

\begin{figure}
\epsscale{1}
\begin{center}
\plotone{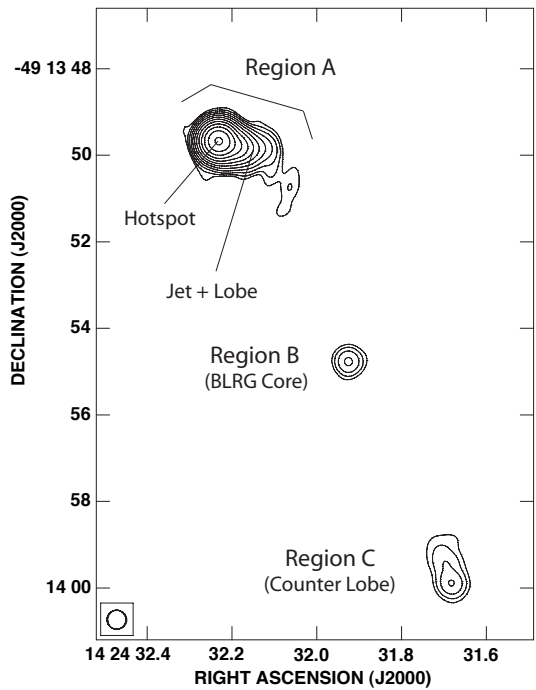}
\end{center}
\caption{ATCA 20.2\GHz image of PKS~1421--490 with source components
  labeled. This image was first presented in G05. To avoid confusion, the
  naming convention used by G05 is also included. Contour levels:
  1.5$\>$mJy/beam $\times$ (1, 2, 4, 8, 16, 32, 64, 128, 256, 512, 1024).
  Peak surface brightness: 2.11\Jy/beam. Beam FWHM: 0.54 $\times$ 0.36
  arcsec. The scale of the image is 7.0 kpc/arcsecond.  \label{fig:ATCA}} 
\end{figure}

Figure \ref{fig:ATCA} illustrates the arcsecond scale radio structure of
PKS~1421--490; it is annotated to show the naming convention for different
components in the radio image used by G05, as well as the correct
interpretation of each of these components brought out in this study. G05
obtained an optical spectrum of region B, and suggested it was not associated
with an active galactic nucleus (AGN) in view of the apparent lack of spectral
lines (due to poor signal to noise ratio in that spectrum). Region A was known
to contain bright VLBI scale radio structure \citep{preston89} and had a flat
radio spectrum ($\alpha < 0.5$). Region B was also known to be much weaker
than region A at radio wavelengths.  Consequently region A was thought to be
an AGN, while  region B was (erroneously) interpreted by G05 as a jet knot. In
this paper we show that in fact region B \textit{is} the active galactic
nucleus (see \S \ref{sec:region_B}), and that region A contains a high surface
brightness hotspot.  The main focus of this paper is the interpretation and
modeling of the exceptional hotspot in region A which has until
now been interpreted as an AGN.

One of the major results of this paper relates to an observed low frequency
flattening in the hotspot radio spectrum at GHz frequencies, which indicates
that the underlying electron energy distribution flattens towards lower
energies. The low energy electron distribution is not only important for calculating parameters such as the number density and energy density, it also provides important constraints on
the particle acceleration mechanism. In turn, this will help to address more
fundamental issues such as jet composition and speed. We now present a brief
overview of the literature relating to the low energy electron distribution in
jets and hotspots.  

In a small number of objects, flattening of the hotspot radio spectra towards
lower frequencies has been observed. In most cases, synchrotron self
absorption and free-free absorption can be ruled out, and the observed
flattening is interpreted in terms of a turn-over in the electron energy
distribution \citep{leahy89, carilli88, carilli91, lazio06}. When modeling
hotspot spectra, the turn-over in the electron energy distribution is usually
approximated by setting the electron number density equal to zero below a
cut-off Lorentz factor $\gamma_{\rm{min}}$. In each case where flattening of
the hotspot radio spectrum has been directly observed, estimates of
$\gamma_{\rm{min}}$ are of the order of several hundred: Cygnus A,
$\gamma_{\rm{min}} \sim$ 300 - 400 \citep{carilli91, lazio06, hardcastle01b};
3C295, $\gamma_{\rm{min}} \sim$ 800 \citep{harris00, hardcastle01b}; 3C123,
$\gamma_{\rm{min}} \sim$ 1000 \citep{hardcastle01a, hardcastle01b};
PKS~1421--490, $\gamma_{\rm{min}} \sim$ 650 (this work).  

\citet{leahy89} presented evidence for a low energy cut-off in two other
hotspots; 3C268.1 and 3C68.1. In both these ojects, the hotspot radio spectra
are significantly flatter between 150MHz and 1.5GHz than they are above
1.5GHz, which suggests a similar value of $\gamma_{\rm{min}}$ to those listed
above, provided the hotspot magnetic field strengths are similar. More
recently \citet{hardcastle01b} reported on a possible detection of an optical
inverse Compton hotspot in the quasar 3C196. By modeling the synchrotron self
Compton emission and assuming a magnetic field strength close to the
equipartition value, they inferred a cut-off Lorentz factor $\gamma_{\rm{min}}
\sim 500$. All of the above listed $\gamma_{\rm{min}}$ estimates appear to be
distributed around a value of $\gamma_{\rm{min}} \sim 600$. However,
\citet{blundell06} and \citet{erlund08} have inferred the existence of a low
energy cut-off at $\gamma_{\rm{min}} \gtrsim 10^4$ in the hotspots of the
giant radio galaxy 6C~0905--3955. Their method of detecting the low energy
cut-off is quite different to those described above, and is based on the
interpretation of an absence of X-ray emission from the eastern hotspot and
radio lobe in that source.  

In \S \ref{sec:gamma_min} we show that a turn-over in the electron energy
distribution at $\gamma_{\rm{min}} \sim 100 - 1000$ can arise naturally from
dissipation of jet energy if the jet has a high proton fraction and a bulk
Lorentz factor $\Gamma_{\rm{jet}} \gtrsim 5$. However, \citet{stawarz07}
have suggested that the low frequency flattening in the radio spectrum of a
Cygnus A hotspot is not related to the turn-over in electron energy
distribution. Rather, they argue, it indicates a transition between two
different acceleration mechanisms.  

Electron energy distributions with a low energy cut-off have also been
discussed in relation to pc-scale jets. The absence of significant Faraday
depolarization in compact sources suggests that the number density of
electrons with Lorentz factor $\gamma \gtrsim 100$ greatly exceeds that of
lower energy particles \citep{wardle77, jones77}. \citet{gopal-krishna04} have
argued that some statistical trends in superluminal pc-scale jets may be
understood in terms of effects arising from a low energy cut-off in the
electron energy distribution. \citet{tsang07} have suggested that a low energy
cut-off in the electron spectrum can alleviate several theoretical
difficulties associated with the inverse Compton catastrophe in compact radio
sources, including anomalously high brightness temperatures and the apparent
lack of clustering of powerful sources at 10$^{12}$ K. However, circular
polarization in the pc-scale jet of 3C279 requires a minimum Lorentz factor
$\gamma_{\rm{min}} < 20$ \citep{wardle98}.  

Observational constraints on the low energy electron distribution in
extragalactic jets on kpc-scales are rare. A low energy cut-off in the
electron energy distribution at $\gamma_{\rm{min}} \sim 20$ has been estimated
for the jet of PKS~0637--752 (500 kpc from the nucleus) through modeling of
the radio to X-ray spectral energy distribution in terms of inverse Compton
scattering of the cosmic microwave background \citep{tavecchio00, uchiyama05}.  

This paper is structured as follows:  In \S \ref{sec:observations} we discuss
our observations and data reduction. In \S \ref{sec:region_B} we discuss the
active galactic nucleus - in particular the optical spectrum and broad band
spectral energy distribution. In \S \ref{sec:hotspot} we present the VLBI
image of the northern hotspot and derive plasma parameters by modeling the
broad band spectral energy distribution. In \S \ref{sec:whole_source_fit} we
independently estimate the hotspot plasma parameters by modeling the radio
spectrum of the entire radio galaxy. In \S \ref{sec:break_frequency} we
discuss the incompatibility of the observed spectrum with the standard
continuous injection plus synchrotron cooling model for hotspots. In \S
\ref{sec:Doppler} we consider Doppler beaming as a possible cause of the
high radio surface brightness and various other properties of the
hotspot. In \S \ref{sec:gamma_min} we consider the dissipation of energy associated with a cold proton/electron jet and present an expression that
relates the energy of the peak in the electron energy distribution to the jet bulk Lorentz factor. We then
consider the implications of this expression in the case of the northern
hotspot of PKS~1421--490 and other objects. In \S \ref{sec:conclusions} we summarize our
findings.  

Throughout this paper we assume cosmology $\Omega_{\Lambda} = 0.73, \Omega_{M}
= 0.27, H_0 = 71 \> \rm{km} \> s^{-1} \> \rm{Mpc}^{-1}$, and we define the
spectral index as $\alpha = -\frac{\rm{log}(F_1/F_2)}{\rm{log}(\nu_1/\nu_2)}$
so that the flux density $F_{\nu} \propto \nu^{-\alpha}$.

\section{Observations and Data Reduction} \label{sec:observations}

\subsection{Summary}

\begin{figure}
\epsscale{1}
\begin{center}
\plotone{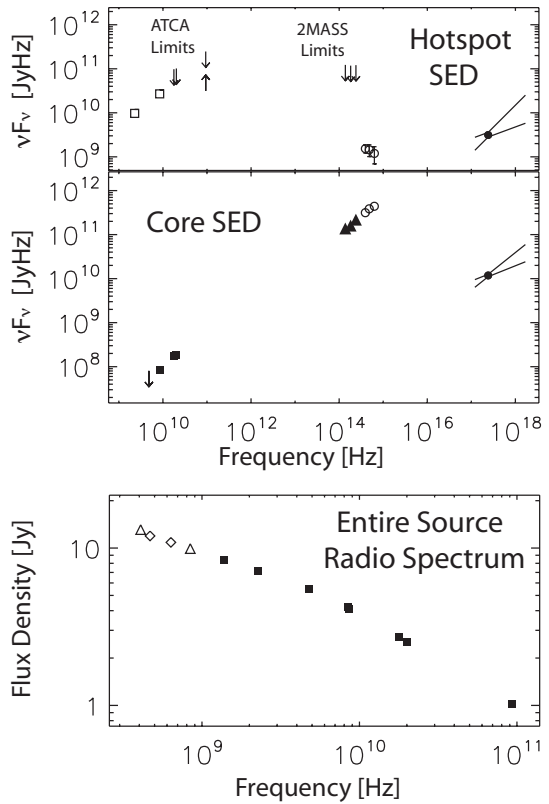}
\end{center}
\caption{Spectral energy distribution of the core and Northern hotspot of PKS1421-490, as well as the radio spectrum of the entire radio galaxy. Symbols indicate the source of the data as follows: Filled squares, ATCA; open squares, LBA; open circles, Magellan (taken from G05); filled triangles, 2MASS (taken from G05); open triangles, MOST (taken from \citet{large81} and \citet{campbell-wilson94}), open diamonds, Parkes (taken from \citet{wills75}); filled circles, Chandra (taken from G05). The solid lines through the filled circles indicate the 1$\sigma$ range of X-ray spectral index permitted by the Chandra data. Tips of arrows mark upper and lower limits (see section \ref{sec:radio_constraints} for discussion of methods used to obtain ATCA limits). Symbol sizes are greater than or approximately equal to error bars except for those points where error bars have been plotted. \label{fig:multiwavelength_plots}} 
\end{figure}

We observed PKS1421-490 with the Long Baseline Array (LBA) at 2.3 and 8.4~GHz and with the ATCA at 2.3, 4.8, 8.4 and 93.5~GHz. We have also made use of ATCA radio data (4.8, 8.6, 17.7 and 20.2~GHz) previously published in \citet{gelbord05} as well as archival 1.4~GHz ATCA data. We combined these data with previously published infra-red, optical and X-ray flux densities to construct radio to X-ray spectra for the northern hotspot and the core as well as an accurate radio spectrum of the entire radio galaxy. Table \ref{table:obs_info} lists the observation information and references for all data used in this study. Figure \ref{fig:multiwavelength_plots} presents the spectra and indicates the source of each data point. 

In addition to these data, we obtained an optical spectrum of region B in order to confirm the classification of that region as an active galactic nucleus. The spectroscopic observations are described in \S \ref{sec:optical_spectroscopy}. 

As well as describing the observations and data reduction steps, this section includes a description of some non-standard procedures that were required to construct the hotspot radio spectrum. Specifically, non-standard procedures were required to determine the hotspot flux density from the 8.4GHz LBA data-set due to limited (u, v) coverage. These non-standard procedures are described in \S \ref{sec:8.4GHz_LBA_flux_determination}. Non-standard procedures were also used to obtain a lower limit to the hotspot flux density at 93.5GHz. This procedure is described in \S \ref{sec:radio_constraints}.

\begin{turnpage}
\begin{deluxetable*}{cccccccc}
\tabletypesize{\scriptsize}
\tablecaption{Observation Information and Flux Densities \label{table:obs_info}}
\tablewidth{0pt}
\tablehead{
\colhead{Flux Density of...} & \colhead{Instrument} & \colhead{Frequency} &
\colhead{Date Observed} & \colhead{Configuration} & \colhead{Resolution \tablenotemark{a}} & \colhead{Flux Density} & 
\colhead{Reference}  \\
&&&&&&[Jy]& \\
}
\startdata
Entire Source &MOST&408\MHz& 1968 - 1978 & --- & 2\farcm8 & 13.1 $\pm$ 0.7& 1\\
Entire Source &Parkes&468\MHz& 1965 - 1969 & --- & 54\arcmin  & 11.9 $\pm$ 0.1& 2\\
Entire Source &Parkes&635\MHz& 1965 - 1969 & --- & 30\farcm5 & 10.9 $\pm$ 0.5& 2\\
Entire Source &MOST&843\MHz& 1990 - 1993 & --- & 1\farcm1  &  9.9 $\pm$ 0.5& 3\\
Entire Source &ATCA & 1.38\GHz & Feb 24 2000 & 6A\tablenotemark{b} &2\farcm2 \tablenotemark{c} &  8.5 $\pm$0.2 & 4 \\
Entire Source &ATCA & 2.28\GHz & Mar 23 2006 & 6C\tablenotemark{b} & 3\arcmin \tablenotemark{c} &  7.15$\pm$0.15 & 4 \\
Entire Source &ATCA & 4.80\GHz &  May 19 2005 & H168 & 3\farcm5 \tablenotemark{c} &  5.5$\pm$0.1 & 4 \\
Entire Source &ATCA & 8.425\GHz & May 19 2005 & H168 & 2\arcmin \tablenotemark{c} & 4.25$\pm$0.1& 4 \\
Entire Source &ATCA & 8.64\GHz & Feb 4 2002 & 6C\tablenotemark{b} & 47\arcsec \tablenotemark{c} &  4.1$\pm$0.1& 4 \\
Entire Source &ATCA & 17.73\GHz & May 9 2004 & 6C\tablenotemark{b} & 30\arcsec \tablenotemark{c} & 2.74$\pm$0.06& 4 \\
Entire Source &ATCA & 20.16\GHz & May 9 2004 & 6C\tablenotemark{b} & 26\arcsec \tablenotemark{c} &  2.54$\pm$0.05& 4 \\
Entire Source &ATCA & 93.5\GHz & Aug 21 2005 & H214\tablenotemark{b} & 10\arcsec \tablenotemark{c} &  1.0$\pm$0.1 & 4 \\
Northern Hotspot &LBA & 2.28\GHz & Mar 23 2006 &  Tidbinbilla, ATCA & $13.5 \times 11.6\> \rm{mas}$ &  4.25 $\pm$ 0.2 & 4 \\
&& & & Mopra, Parkes  & &  &  \\
&& & & Hobart, Ceduna  & &  &  \\
Northern Hotspot&LBA & 8.425\GHz & May 19 2005 &  Parkes, Mopra, ATCA & $33 \times 13 \> \rm{mas}$ &  3.2 $^{+0.2}_{-0.3}$ & 4 \\
Northern Hotspot& ATCA & 17.73\GHz & May 9 2004 & 6C & 0\farcs58 $\times$ 0\farcs43 &  $<$ 2.3 & 4 \\
Northern Hotspot& ATCA & 20.16\GHz & May 9 2004 & 6C & 0\farcs51 $\times$ 0\farcs37  &  $<$ 2.1 & 4 \\
Northern Hotspot& ATCA & 93.5\GHz & Aug 21 2005 & H214 & 10\arcsec \tablenotemark{c} &  0.8 $ < F_{93.5 \> \rm{GHz}} < $  1.1\tablenotemark{d} & 4 \\
Northern Hotspot& 2MASS & $1.38 \times 10^{14}$~Hz & 1998 - 2001 & --- & $\sim$4\arcsec  &  $< 3.7 \times 10^{-4}$ & 5 \\
Northern Hotspot& 2MASS & $1.82 \times 10^{14}$~Hz & 1998 - 2001 & --- & $\sim$4\arcsec  &  $< 2.7 \times 10^{-4}$ & 5 \\
Northern Hotspot& 2MASS & $2.4 \times 10^{14}$~Hz & 1998 - 2001 & --- & $\sim$4\arcsec  &  $< 2.1 \times 10^{-4}$ & 5 \\
Northern Hotspot& Magellan & $3.93 \times 10^{14}$~Hz & Apr 26 2003 & MagIC &  $\sim$0\farcs6 &  $(3.9 \pm 0.8) \times 10^{-6}$ & 5 \\
Northern Hotspot& Magellan & $4.82 \times 10^{14}$~Hz & Apr 26 2003 & MagIC &  $\sim$0\farcs6 &  $(3.0 \pm 0.9) \times 10^{-6}$ & 5 \\
Northern Hotspot& Magellan & $6.29 \times 10^{14}$~Hz & Apr 26 2003 & MagIC &  $\sim$0\farcs6 &  $(1.9 \pm 0.8) \times 10^{-6}$ & 5 \\
Northern Hotspot& Chandra & $2.41 \times 10^{17}$~Hz & Jan 16 2004 & ACIS-S & 0\farcs5 &  $(1.3 \pm 0.16) \times 10^{-8}$ & 5 \\
Core & ATCA & 4.8\GHz & Feb 4 2002 & 6C & &  $< 7 \times 10^{-3}$ & 5 \\
Core & ATCA & 8.64\GHz & Feb 4 2002 & 6C & &  $(9.6 \pm 0.6) \times 10^{-3}$ & 5 \\
Core & ATCA & 17.73\GHz & May 9 2004 & 6C & &  $(9.8 \pm 0.3) \times 10^{-3}$ & 5 \\
Core & ATCA & 20.16\GHz & May 9 2004 & 6C & &  $(9.2 \pm 0.2) \times 10^{-3}$ & 5 \\
Core & 2MASS & $1.38 \times 10^{14}$~Hz & 1998 - 2001 & --- & $\sim$4\arcsec  &  $(1.00 \pm 0.07) \times 10^{-3}$ & 5 \\
Core & 2MASS & $1.82 \times 10^{14}$~Hz & 1998 - 2001 & --- &  $\sim$4\arcsec &  $(8.8 \pm 0.6) \times 10^{-4}$ & 5 \\
Core & 2MASS & $2.4 \times 10^{14}$~Hz & 1998 - 2001 & --- &  $\sim$4\arcsec  &  $(9.1 \pm 0.7) \times 10^{-4}$ & 5 \\
Core & Magellan & $3.93 \times 10^{14}$~Hz & Apr 26 2003 & MagIC & $\sim$0\farcs6  &  $(8 \pm 1) \times 10^{-4}$ & 5 \\
Core & Magellan & $4.82 \times 10^{14}$~Hz & Apr 26 2003 & MagIC & $\sim$0\farcs6  &  $(8 \pm 1.5) \times 10^{-4}$ & 5 \\
Core & Magellan & $6.29 \times 10^{14}$~Hz & Apr 26 2003 & MagIC & $\sim$0\farcs6  &  $(7 \pm 2) \times 10^{-4}$ & 5 \\
Core & Chandra & $2.41 \times 10^{17}$~Hz & Jan 16 2004 & ACIS-S & 0\farcs5  &  $(4.9 \pm 0.3) \times 10^{-8}$ & 5 \\
\enddata
\tablecomments{The uncertainties in ATCA flux density are dominated by the
  uncertainty in the absolute flux calibration, which is estimated to be
  2$\%$, except for 93.5$\>$GHz, where the uncertainty is estimated to be
  10$\%$. The lower limit on the hotspot flux at 93.5\GHz comes from making an
  assumption about the non-hotspot spectrum extrapolated to higher frequencies
  from 8.4$\>$GHz. See section \ref{sec:radio_constraints} for
  details. References: (1) \citet{large81}, (2) \citet{wills75}, (3)
  \citet{campbell-wilson94}, (4) This work, (5) \citet{gelbord05}. } 
  \tablenotetext{a}{Convert to linear resolution using 7.0~kpc/arcsecond}
\tablenotetext{b}{Only short baselines on which the radio galaxy is unresolved
  were used to measure the flux density.} 
  \tablenotetext{c}{Resolution of shortest baseline.}
     \tablenotetext{d}{See section \ref{sec:radio_constraints}.}
\end{deluxetable*}
\end{turnpage}

\subsection{VLBI}

\subsubsection{LBA Observations at 2.3\GHz}

PKS~1421--490 was observed with 6 elements (ATCA, Mopra, Parkes, Tidbinbilla
70m, Hobart and Ceduna) of the Australian Long Baseline Array (LBA) on March
23 2006. A full 12 hour synthesis was obtained, recording a single 16\MHz
bandwidth in both left and right hand circular polarization. Regular scans on
a nearby phase calibrator, PKS~1424--418, were 
scheduled throughout the observation, as well as scans on a point like source,
PKS~1519--273 \citep{linfield89}, used for gain calibration. Unfortunately,
due to hardware issues, we 
were only able to process right hand circular polarization. However, this will not affect our results as we do not expect the hotspot emission to be significantly circularly polarized. Circular polarization in AGN seldom exceeds a few tenths of 1$\%$ \citep{rayner00}. Data were recorded
to VHS tapes using the S2 system and correlated using the LBA hardware
correlator with 32 channels 
and 2 second integration time. The data were correlated twice: once with the
phase tracking centre located at the position of the radio peak in region A, and once with the
phase tracking centre located $\sim 5$ arcseconds away, at the position of
the core (region B). 

The initial calibration of the visibility amplitudes was performed in AIPS, using
the measured system temperatures and antenna gains. We obtained simultaneous ATCA data during our
observation, and this allowed us to bootstrap the LBA flux scale to the ATCA
flux scale by comparing simultaneous measurements of the point like source PKS~1519--273. After scaling the gains using this bootstrapping method, and correcting the residual delays and rates via fringe fitting, the data-set from the phase reference source was exported to DIFMAP \citep{shepherd97} where
it was edited and imaged. Amplitude and phase self calibration corrections
obtained from imaging the phase reference source were imported into AIPS using the
\emph{cordump}\footnote{The \emph{cordump} patch is available for DIFMAP at \url{http://astronomy.swin.edu.au/$\sim$elenc/DifmapPatches/}} patch kindly supplied to us by Emil Lenc. These phase and amplitude
corrections were then applied to PKS~1421--490, and the data exported to DIFMAP for deconvolution and self calibration. The resulting image is shown in figure 
\ref{fig:LBA}. We measure a hotspot flux density of $4.25 \pm 0.2 \>$Jy at 2.3$\>$GHz.  Preston et al. (1989) obtained a flux density of 4.1~Jy at 2.3~GHz for the northern hotspot by model fitting SHEVE (Southern Hemisphere VLBI Experiment) data with the simplest model consistent with the data (two circular Gaussians).

We attempted to detect compact structure within region B (the core) using the
data-set that had been correlated with the phase centre at that position. The
time-averaging- and bandwidth-smeared emission from region A in this dataset
was first cleaned to remove the side-lobes, but we were unable to detect any
emission from the location of the core to a limit of approximately 8$\>$mJy (5
$\sigma$). The upper limit to the flux density of region B at 4.8$\>$GHz is
7$\>$mJy \citep{gelbord05} 

\subsubsection{LBA Observations at 8.4\GHz}

PKS~1421--490 was observed with 5 elements of the LBA (ATCA, Mopra, Parkes,
Hobart and Ceduna) in May 2005 at 8.4$\>$GHz. A full 12 hour synthesis was 
obtained, recording a single 16\MHz bandwidth in both left and right hand
circular polarization. At this frequency, the Northern hotspot of
PKS~1421--490 is completely resolved on all but the shortest three baselines
(baselines between Parkes, the ATCA and Mopra). We fringe fitted our target
source data using a point source model in AIPS, before performing
model-fitting  and phase self calibration iterations in DIFMAP.  

\subsubsection{Determination of Hotspot Flux Density from the 8.4\GHz LBA Data Set} \label{sec:8.4GHz_LBA_flux_determination}

Due to the small number of baselines, we use model-fitting in the
(\emph{u, v})-plane rather than CLEAN deconvolution to measure the hotspot
flux density at 8.4$\>$GHz. Model-fitting involves specifying a starting model
in the image plane, consisting of a number of elliptical Gaussian components,
each with a particular flux density, position, size and position angle, then
allowing the model fitting algorithm to locate a chi-squared minimum by
fitting the Fourier transform of the model to the (\emph{u, v})-data.  

Care is necessary when comparing the LBA flux density measurements at the two
different frequencies, due to the limited (\emph{u, v}) coverage. At
2.3$\>$GHz, the data cover (\emph{u, v}) spacings between 0.5 and 13
M$\lambda$, while at 8.4\GHz the data cover (\emph{u, v}) spacings between 2
and 9 M$\lambda$. Therefore, provided the source structure can be described by
a simple model consisting of a set of Gaussian components, the comparison of
model flux densities will be valid. 

In order to determine
the range of allowable flux densities in the 8.4~GHz data-set, we specified a wide range of different
models consisting of 3, 4 or 5 elliptical Gaussian components, broadly
consistent with the 2.3~GHz image, then let the model-fitting algorithm adjust
the model to fit the 8.4~GHz data. While it is not possible to precisely
constrain the flux density of the hotspot with only three baselines, we found
that the total flux density of all acceptable models (using between three and
five elliptical Gaussian components and a wide range of initial model
parameters) was never less than 2.9$\>$Jy, and the flux density of the
best-fitting model was 3.2$\>$Jy. The 8.6\GHz ATCA image contains an
unresolved source of 3.3~Jy at the position of the hotspot, and this provides
an upper limit to the hotspot flux density at 8.6$\>$GHz. We therefore adopt a hotspot flux density at 8.4~GHz of $F_{8.4 \> \rm{GHz}} = 3.2^{+0.2}_{-0.3}
\>$Jy.  

\vspace{0.5cm}

\subsection{ATCA Observations}

ATCA observations of PKS~1421--490 were made simultaneously during our LBA
observations. We recorded a single 64\MHz bandwidth at 8.4\GHz and 128\MHz
bandwidth at 4.8\GHz during our first LBA observation in 2005. A single
128\MHz bandwidth at 2.3\GHz was recorded during our second LBA observation in
2006. For each of 
these observations, the ATCA was
in a compact configuration so we could not image the source in detail, but we
were able to obtain accurate total source flux density measurements (see Table
\ref{table:obs_info}).  
Total source flux density measurements were also obtained at 1.4\GHz using
archival ATCA data. Standard calibration and imaging procedures were used 
with the MIRIAD processing software. 

In August 2005
we obtained a full 12 hour synthesis with the new
3mm receivers. Again, standard calibration and imaging procedures were used in
MIRIAD. The flux density scale was determined from scans on the planet Uranus
and confirmed using the point-like source PKS~1921--293, the flux density of
which had been measured 4 days prior to our observing run as $8.8 \pm 0.9
\>$Jy at 93.5$\>$GHz. We detected a single point like component in the 93.5GHz
image of PKS~1421--490 coincident with region A, the flux density of which we
regard as being the total source flux density at this frequency, since the
resolution of the shortest baseline larger than the source. The
upper limit on flux density at 93.5$\>$GHz for region B and C is 5$\>$mJy (5
$\sigma$). 

Errors in the flux densities reported in Table \ref{table:obs_info} are
dominated by uncertainties in the primary flux calibration which are estimated
to be of order $\sim$2$\%$ at cm wavelengths, and of order 
$\sim$10$\%$ at 3mm. 

\subsection{Constraints on the Hotspot Radio Spectrum from ATCA Images} \label{sec:radio_constraints}

The hotspot is unresolved in the ATCA images, and is blended with emission from the surrounding regions. We are therefore unable to directly measure the flux density of the hotspot from the ATCA data. However, we are able to constrain the flux density of the hotspot, and we now discuss the methods used to obtain upper and lower limits. 

Using radio data that was first presented in \citet{gelbord05}, we find
upper limits on the hotspot flux density at 17.7\GHz and 20.2\GHz by summing
the CLEAN components at the position of the hotspot. Similarly, we obtain an upper limit to the hotspot flux density at 93.5\GHz from
the measurement of total source flux density at that frequency. We obtain a
lower limit on the hotspot flux density at 93.5~GHz via the following steps: 

\begin{itemize}
\item[1.] We subtract the LBA-measured hotspot flux density from the total
  source flux density at 2.3\GHz and 8.4\GHz to obtain two estimates of the
  non-hotspot flux density, from which we calculate a non-hotspot spectral
  index ($\alpha_{2.3 \> \rm{GHz}}^{8.4 \> \rm{GHz}}$(non-hotspot) = 0.78). 
\item[2.] We extrapolate the non-hotspot power law to 93.5$\>$GHz.
\item[3.] We reasonably assume that the non-hotspot spectrum becomes steeper
  towards higher frequencies. Therefore the extrapolated flux density from step 2 is an upper limit to the non-hotspot flux density.   
\item[4.] We subtract the non-hotspot upper limit from the observed entire
  source flux density to obtain a lower limit on the hotspot flux density at
  93.5$\>$GHz. 
\end{itemize}
The assumption in step 3 is based on the observation that the non-hotspot
emission arises in the lobes, jets and the southern hotspot (the core is
negligible). Jet and lobe spectra are often observed to steepen towards higher
frequency. Indeed, the 17.7\GHz and 20.2\GHz ATCA images indicate that the
spectral index of the northern lobe region steepens significantly at higher frequency. The
limits on hotspot flux density obtained from the ATCA images are represented
by the tips of the arrows in Figure \ref{fig:multiwavelength_plots}, and are listed in Table \ref{table:obs_info}.

\begin{figure}
\epsscale{1.1}
\begin{center}
\plotone{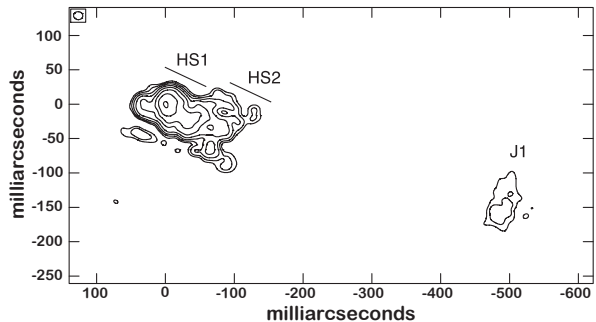}
\caption{VLBI image of the Northern Hotspot of PKS~1421--490 at
  2.3$\>$GHz. Contour levels: 3.5$\>$mJy/beam $\times$ (1, 2, 4, 8, 16, 32,
  64, 128). Peak surface brightness: 0.47$\>$Jy/beam. Beam FWHM: 13.5mas
  $\times$ 11.6mas. The scale of this image is 7.0
  pc/milli-arcsecond. \label{fig:LBA}} 
\end{center}
\end{figure}

\subsection{Optical Spectroscopy} \label{sec:optical_spectroscopy}

Optical spectra were taken with the Magellan IMACS camera on 14 May 2005 in
service mode.  Three ten minute exposures were obtained using a long slit
(0\farcs9 width) aligned with regions A and B.  The 300 lines/mm grism was
used, to yield a spectral resolution of $R \sim 1000$ spanning roughly
4000--10000 \AA.  The spectra were reduced with IRAF.  No standard stars were
observed, so no effort was made to flux calibrate the spectra or to remove
telluric absorption features. 

No significant spectral features were detected in the spectrum of region A,
consistent with synchrotron emission from a hotspot.  However, only a very
high equivalent width emission line could have been detected due to the low
signal to noise ratio of these data.  

The spectrum of region B contains several broad and narrow emission lines,
which allowed a precise determination of the redshift (see \S
\ref{sec:region_B_spectrum}).  This is not the first spectrum of the nucleus to be published --- a spectrum of region B was presented in G05.  However, the spectrum presented in G05 did
not allow identification of any spectral lines because the strongest spectral
features fell beyond the wavelength coverage, and the spectrum was taken as a
single short exposure with a high background due to the pre-dawn sky. 

\vspace{1cm}

\section{Region B: The Active Galactic Nucleus} \label{sec:region_B}

\subsection{Optical Spectrum of Region B} \label{sec:region_B_spectrum}

\begin{figure}
\epsscale{1.1}
\plottwo{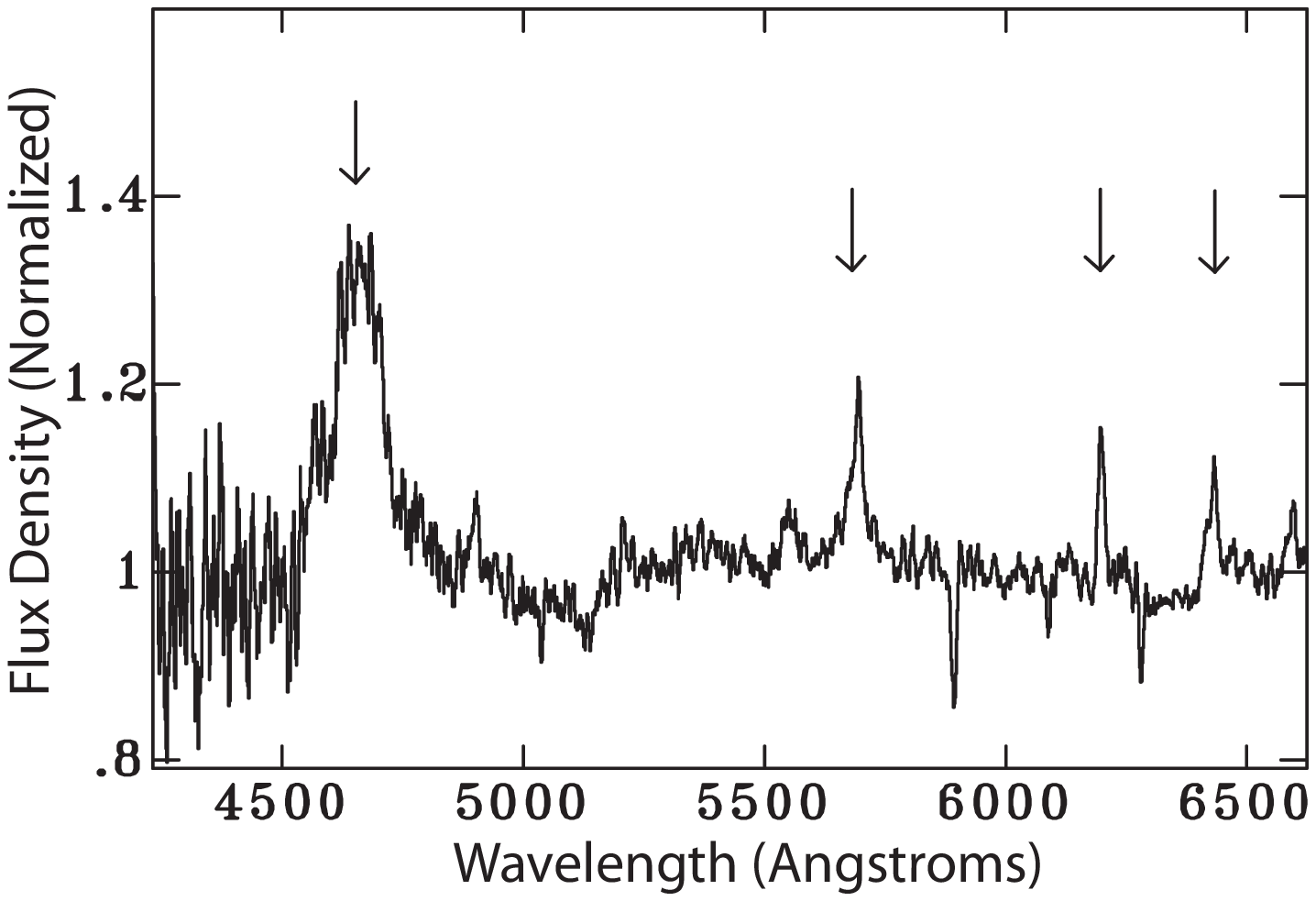}{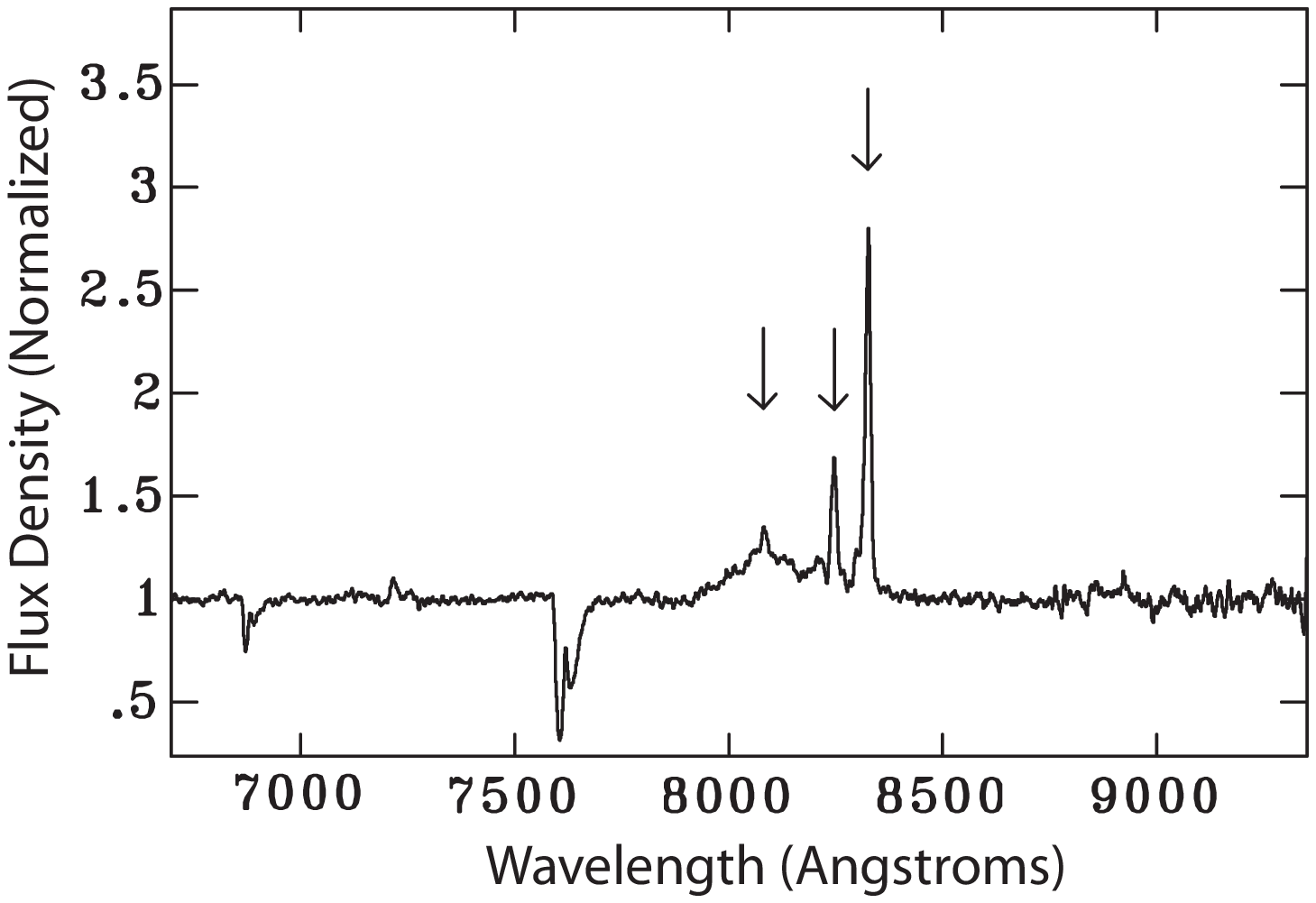}
\caption{The normalized optical spectrum of region B. This spectrum clearly
  shows that region B is a broad line AGN at redshift 0.6628. Increasing in
  wavelength, the arrows indicate the positions of the following emission
  lines: Mg II 2799, [Ne V] 3426, [O II] 3726/3729, [Ne III] 3869, H$\beta$,
  [O III] 4959, [O III] 5007. \label{fig:region_B_spectrum}} 
\end{figure}

The normalized optical spectrum of region B is displayed in Figure
\ref{fig:region_B_spectrum}. We detect several broad and narrow emission
lines: Mg II 2799, [Ne V] 3346 and 3426, the blended [O II] 3726,3729 doublet,
[Ne III] 3869 and 3967, H$\delta$ (marginal), H$\gamma$, [O III] 4363,
H$\beta$, and [O III] 4959 and 5007.  H$\beta$ has both a broad and narrow
component; their measured FWHM values (uncorrected for instrumental
resolution) are 6500 km/s and 516 km/s respectively.  The narrow H$\beta$ line
width is consistent with that of the OIII lines (510 km/s).  From these
features we measure the redshift $z = 0.6628 \pm 0.0001$.

\subsection{Spectral Energy Distribution of the Core}

We did not detect the active galactic nucleus with the LBA, but obtain an
upper limit on the flux density of approximately 8$\>$mJy at 2.3$\>$GHz. This is consistent with the ATCA core flux density measurements (see Table \ref{table:obs_info}). The
core is completely dominated by its optical 
emission. In fact, the optical flux density is so great relative to the radio
($\alpha_{8.6 \> \rm{GHz}}^{662.4 \> \rm{nm}}$ = 0.23 $\pm 0.02$), the core
would be 
classified as radio quiet in the strictest sense. The radio to optical
spectral index cannot be explained in terms of standard synchrotron self
absorption models for flat radio spectra \citep[eg.][]{marscher88}, since this
would require at least part of the jet to be self absorbed at optical
wavelengths and would imply an unrealistically high magnetic field
strength. This region has an optical spectral index $\alpha_o = 0.2 \pm 0.1$,
in the range typical of quasars \citep{francis91}, suggesting that the strong
optical emission may be due to an unusually large contribution from the
accretion disk thermal component. The radio to X-ray spectral index
($\alpha_{8.6 \> \rm{GHz}}^{1 \> \rm{keV}}$ = 0.710 $\pm 0.005$) is typical of
radio galaxies at similar redshift \citep[eg.][]{belsole06}. The optical to
X-ray spectral index is $\alpha_{ox} = 1.62$ (G05). There is clearly an excess
of optical flux relative to the radio and X-ray flux when compared to samples
of other radio galaxies \citep[eg.][]{gambill03}.  Note that measurements of the B-band magnitude have shown no variability, to within 0.6 magnitudes, over the past 35 years (GO5).

\section{The Northern Hotspot} \label{sec:hotspot}

\subsection{Morphology} \label{sec:morphology}

Figure \ref{fig:LBA} shows the LBA image of the Northern
hotspot at 2.3$\>$GHz. Less than 0.2$\>$Jy (5$\%$ of the hotspot flux density)
remains on the longest baseline  ($\sim$ 12.9M$\lambda$), implying that there
is little structure on scales smaller than 15 mas (100 pc). This limit on
substructure within the hotspot is relevant to possible synchrotron self
absorption models for the hotspot spectrum, which we discuss further in \S
\ref{sec:modelling}.

The flux density of the hotspot in our LBA image is 60 $\%$ of the total
source flux density at 2.3$\>$GHz, and 75 $\%$ at 8$\>$GHz.  The peak surface
brightness is $I_{\rm{2.3 \> GHz}}^{\rm{peak}} = 2600 \>
\rm{Jy/arcsec^2}$. Extrapolating to 8$\>$GHz assuming $I_{\nu} \propto
\nu^{-0.2}$ (the spectral index of $\alpha \sim 0.2$ between these frequencies
is calculated in \S \ref{sec:modelling}) and accounting for cosmological dimming and redshifting, we find that the peak surface brightness of the northern hotspot of PKS~1421-490 would be more than 1000 times brighter than the brightest hotspot of Cygnus A if they were at the same redshift \citep{carilli99}. The monochromatic hotspot luminosity is $L_{2.3 \> \rm{GHz}} = 8 \times 10^{27}$ WHz$^{-1}$.  

A protrusion on the Eastern edge of the hotspot resembles the
``compact protrusions'' seen in numerical simulations
\citep[eg.][]{norman96}. According to \citet{norman96}, a compact protrusion
is produced in their 3-D non-relativistic hydrodynamic simulations when the
light, supersonic jet reaches the leading contact discontinuity. At this
point, the jet is generally flattened to a width substantially less than the
inlet jet diameter, and the compact protrusion arises where the jet impinges
on the contact discontinuity surface.  

The width of the hotspot at the peak (region HS1) is 400pc measured
perpendicular to the inferred jet direction. The lower surface brightness
emission behind the hotspot peak (region HS2) is 700pc measured perpendicular
to the jet direction. The length of the hotspot (regions HS1 and HS2) is
approximately 1kpc. The geometric mean of  the major and minor axes (for
comparison with \citet{hardcastle98} and \citet{jeyakumar00}) is 0.63
kpc. This is a factor of 4 below the median value (2.4 kpc) of hotspot sizes
given in \citet{hardcastle98}. However, the size of the hotspot relative to
the linear size of the source is consistent with the correlation between these
parameters given in \citet{hardcastle98} and \citet{jeyakumar00}.  

The jet exhibits a bend of almost 60 degrees (projected) approximately 5
arcseconds (35kpc) from the core at the western end of the ridge of emission
extending west from the hotspot in Figure \ref{fig:ATCA}. \citet{bridle94}
showed that hotspot brightness is anti-correlated with apparent jet deflection
angle. They found that, for the twelve quasars in their sample, the ratio of
hotspot flux density to lobe flux density decreases with larger jet bending
angles, particularly when the deflection occurs abruptly. PKS~1421--490 does
not follow this trend.  

We detect what appears to be a jet knot (region J1) 512 mas ($\sim$ 3.5 kpc
projected) at position angle -107 degrees (North through East) from the
hotspot peak. The knot is extended along a position angle almost perpendicular
to the apparent jet direction. The major axis of the knot is poorly
constrained due to the low signal to noise of this component, but the data
suggests a width of approximately 400 - 600pc.

\subsubsection{Interpretation of Region HS2} \label{sec:region_HS2}

We now consider the interpretation of the lower surface brightness region HS2
just behind the hotspot peak. As mentioned above, the diameter of region HS2 perpendicular to the jet direction (700pc) is much larger than the diameter of region HS1 (400pc). The surface brightness of region HS2 is more than a factor of 10 times the peak surface brightness of the brightest hotspot of Cygnus A. In addition, the flux density from region HS2 alone ($\sim$
0.6~Jy at 2.3$\>$GHz) is more than 4 times the total flux density of the whole
counter lobe and hotspot. There are two possible interpretations for region HS2, and the interpretation of this region has implications for the interpretation of region HS1. 

The first interpretation is in terms of emission from turbulent back-flow in
the cocoon. If this interpretation is correct, we cannot
appeal to Doppler beaming to account for the high surface brightness of
region HS2 relative to other hotspots, and the high flux density relative to the
counter hotspot and lobe. If we cannot appeal to Doppler beaming for region HS2, it would seem unreasonable to appeal to Doppler
beaming to explain the high surface brightness of region HS1. The arm length symmetry places a tight upper limit on the expansion velocity of the lobes at $v_{\rm{expansion}} < 0.1 c$, indicating that the whole complex (region HS1 and HS2) cannot be advancing relativistically. 

The second possible interpretation for region HS2 is that the emission is
associated with oblique shocks in the jet as it approaches the hotspot.  In
this case, we may appeal to Doppler beaming to explain the high surface
brightness of both regions HS1 and HS2. However, this interpretation would
imply that the jet diameter at HS2 ($\sim$700pc) is significantly greater than the diameter of the hotspot at HS1 ($\sim$400pc) and also greater than the jet diameter at J1 ($\sim400 - 600$pc). It should be noted that the width of region J1, presumably associated with a jet knot, is poorly
constrained due to the low signal to noise of this component. Future LBA observations at 1.4GHz may provide better constraints on the size of regions J1 and HS2. 

\subsection{Modeling the Hotspot Spectral Energy Distribution}  \label{sec:modelling} 

\subsubsection{Low Frequency Flattening} \label{sec:low_frequency_flattening}

\begin{figure}
\epsscale{1}
\begin{center}
\plotone{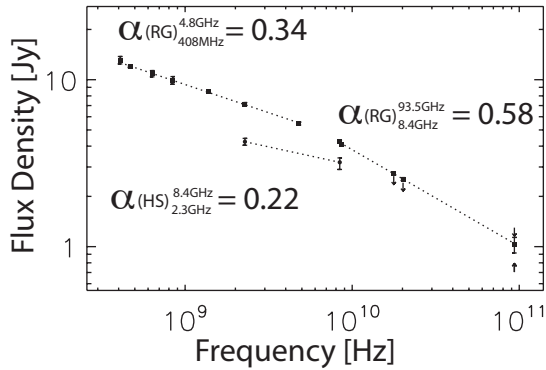}
\end{center}
\caption{Radio spectrum of the entire radio galaxy (filled squares) and hotspot (filled circles and tips of arrows). This figure serves to illustrate the low frequency flattening in the hotspot radio spectrum referred to in the text (\S \ref{sec:low_frequency_flattening} and \S \ref{sec:gamma_min}). The dotted lines illustrate power law fits to specific sections of the data. 
\label{fig:plot_showing_flattening}} 
\end{figure}

Figure \ref{fig:plot_showing_flattening} illustrates that the hotspot radio spectrum changes slope at GHz frequencies, becoming flatter towards lower frequency. We now discuss this
feature in more detail and consider the possible causes.

The hotspot spectral index calculated from our LBA flux density measurements
is relatively flat at GHz frequencies ($\alpha^{8.4 \> \rm{GHz}}_{2.3 \>
  \rm{GHz}} = 0.22_{-0.05}^{+0.08}$). The hotspot spectrum cannot continue
with this slope to millimeter wavelengths, since it would substantially
over-predict the observed 93.5\GHz flux density. We therefore require that the
hotspot spectrum be steeper at frequencies above 8\GHz, with spectral index $\alpha
\gtrsim \alpha^{\rm{ATCA} \; 93.5 \> \rm{GHz}}_{\rm{LBA } \; 8.4 \> \rm{GHz}}
= 0.48$.  

Our conclusion of a flat spectral index at GHz frequencies based on the LBA
flux density measurements is strengthened by inspection of the whole source
spectrum (see Figure \ref{fig:plot_showing_flattening}). The flux density of the
entire radio galaxy has a spectral index of 0.58 above 8.4$\>$GHz, but flattens
to a spectral index of 0.34 below 4.8$\>$GHz. The northern hotspot is the
dominant component at GHz frequencies, hence, the flattening of the total
source spectrum implies there is flattening in the hotspot spectrum. There are
a number of possible causes of this GHz frequency flattening, but most of them are
implausible. We now consider a number of such explanations. 

If synchrotron self absorption were responsible for the flattening, the
required magnetic field strength is $B_{\rm{G}} \sim 10^{-5} \nu_p^5 \theta^4
F^{-2}_p (1+z)^{-1} $ where $B_{ \rm{G}}$ is the magnetic field strength in
Gauss, $F_p$ is the peak flux density in Jy, $\nu_p$ is the frequency in GHz
at the peak and $\theta$ is the angular size in milliarcseconds
\citep[][pg. 325]{deyoung02}. In the case of the Northern hotspot of
PKS~1421--490 we estimate (conservatively) $\nu_p \sim 1 \> \rm{GHz}$, $\theta
\sim 100 \> \rm{mas}$, $F_p \sim 5 \> \rm{Jy}$ and z = 0.663. Therefore, a
magnetic field strength of $B \sim 20 \> \rm{G}$ is required to produce the
observed flattening ---  four orders of magnitude greater than the
equipartition magnetic field strength. Less than 0.2Jy (5$\%$ of the hotspot
flux density) remains on the longest baselines (12.9 M$\lambda$ $\Rightarrow$
15~mas resolution), implying that the hotspot cannot be composed of many small
self-absorbed sub-components. Therefore, we do not consider synchrotron self
absorption to be a viable explanation for the flattening.  

We next consider free-free absorption by interstellar clouds in the hotspot environment as a possible mechanism for the observed flattening of the radio
spectrum. Consider a cloud of size $L_{\rm{kpc}} \> \rm{kpc}$, temperature
T$_4 \times 10^4$ K, electron number density $n_e \> \rm{cm}^{-3}$ and
pressure $p_{-12}\times 10^{-12} \> \rm{dyn/cm^2}$. The optical depth $\tau$
to free-free absorption at a frequency $\nu_{\rm{GHz}} \> \rm{GHz}$ is given
by \citep[eg.][pg.326]{deyoung02} 
\begin{eqnarray}
\tau &=& 3.3 \times 10^{-4} n_e^2 L_{\rm{kpc}} \nu^{-2.1}_{\rm{GHz}} T_4^{-1.35} \\
&=& 2 \times 10^{-4} L_{\rm{kpc}} \; \nu^{-2.1}_{\rm{GHz}} \; p^2_{-12} \;
\left( \frac{n_e}{n} \right)^2 T_4^{-3.35}  
\end{eqnarray}
Assuming a characteristic pressure in the outer regions of an elliptical
galaxy $p \sim 10^{-12} \> \rm{dyn/cm^2}$, a characteristic temperature for an
ionized cloud $T \sim 10^4 \> \rm{K}$ and a reasonable cloud size $L \lesssim
1 \> \rm{kpc}$, the optical depth to free-free absorption above 1\GHz is less
than $5 \times 10^{-5}$.   

We therefore interpret the change of slope in the hotspot radio spectrum in
terms of a change in the underlying electron energy distribution. In \S
\ref{sec:ssc_modeling} we model the SED by incorporating a low energy cut-off
in the electron energy distribution. A low-energy cut-off at some
minimum Lorentz factor $\gamma_{\rm{min}}$ produces a spectrum with $F_{\nu}
\propto \nu^{1/3}$ at frequencies below the characteristic emission frequency
of electrons with Lorentz factor $\gamma_{\rm{min}}$ \citep[see
  eg.][]{worrall06}. We must emphasize that an instantaneous
cut-off in number density is not physical - it is merely an approximation to a
sharp turn-over in the electron energy distribution. In \S \ref{sec:gamma_min} we show that interpreting the observed flattening in terms of a turn-over in the electron energy distribution has considerable implications.  

Low frequency flattening in hotspot spectra has been observed in a small
number of other objects (see \S
\ref{sec:intro}).  

\subsubsection{The High Frequency Synchrotron Spectrum}

The hotspot spectrum remains relatively flat between 8GHz and 93.5GHz, having spectral index $0.4 < \alpha < 0.6$ (based on the two point spectral index from the 8.4GHz LBA data point to the ATCA upper and lower limits at 93.5GHz). The synchrotron spectrum above 93 GHz is poorly constrained, but the simplest model --- a power law spectrum with spectral index $0.4 < \alpha < 0.6$ and an exponential cut-off at high frequency (i.e. a synchrotron spectrum from a power law electron energy distribution with number density set to zero above $\gamma_{\rm max}$), is unable to satisfy the optical data and the 2MASS infra-red upper limits simultaneously. Either a break to a steeper spectrum
somewhere between $\sim 10^{11} - 10^{13} \>$ Hz is required, or a gradual
cut-off at high electron energy, rather than an abrupt cut-off at
$\gamma_{\rm{max}}$, must exist. Given the high radio luminosity, hence
high magnetic field strength, synchrotron losses are likely to be
important. We therefore allow for a synchrotron cooling break at an arbitrary
break frequency in order to fit the radio through optical spectrum. Different choices of model spectrum are possible, but they would not significantly affect our major results. We discuss the model electron energy distribution in \S \ref{sec:ssc_modeling}, and further discuss the self-consistency of this model in \S \ref{sec:break_frequency}.

\subsubsection{Hotspot X-ray Emission} 		
\begin{figure}
\begin{center}
\plotone{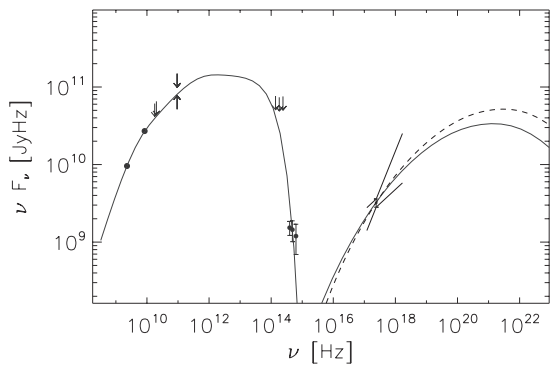}
\caption{Hotspot Spectral Energy Distribution with SSC model overlaid. The
  ``bow-tie'' around the X-ray data point indicates the 1$\sigma$ range of
  X-ray slopes. The solid line is the best fit synchrotron plus self Compton
  model spectrum with the Doppler factor fixed at $\delta = 1$. The dashed
  line is the best fit SSC component with the Doppler factor fixed at
  $\delta=3$. The model synchrotron spectra for $\delta= 3$ and $\delta=1$ are
  exactly the same, and so do not appear as separate curves in the plot. The
  LBA data points (plotted as filled circles) have error bars smaller than the
  symbol size. Tips of arrows mark the position of upper and lower
  limits. X-ray, optical and infra-red points are taken from
  \citet{gelbord05}.  \label{fig:SED}} 
\end{center}
\end{figure}

Figure \ref{fig:SED} shows the spectral energy distribution (SED) of the
northern hotspot. The level of X-ray flux density relative to the optical flux
density indicates the presence of two distinct spectral components:
synchrotron emission from radio to optical frequencies, and inverse Compton
emission at X-ray frequencies and above.

The energy density of the locally generated synchrotron emission within the
hotspot  (assuming no Doppler beaming) is more than $10^4$ times the energy
density of the cosmic microwave background (CMB) at this redshift. If the
hotspot plasma is moving relativistically with velocity $v = \beta c$ at an
angle $\theta$ to the line of sight, the ratio of synchrotron to CMB energy
density is reduced by a factor of  $\sim \Gamma^{-2} \delta^{-3}$ where
$\delta = [\Gamma (1 - \beta \cos \theta)]^{-1}$ is the Doppler factor and
$\Gamma = (1 - \beta^2)^{-1/2}$ is the bulk Lorentz factor, and we have
assumed the hotspot is associated with plasma moving through a stationary
volume/pattern rather than a moving blob, so that $I_{\nu} = \delta^2
I^{\prime}_{\nu^{\prime}}$ \citep{lind85}. Therefore, if $\delta \lesssim 6$
(assuming $\Gamma \sim \delta$), inverse Compton scattering of locally
generated synchrotron photons is the dominant source of inverse Compton X-ray
emission. While a Lorentz factor of $\Gamma \sim 6$ is not ruled out, such a
high Lorentz factor is not required by the data, and we consider only Lorentz
factors $\Gamma \lesssim 3$. We therefore ignore the inverse Compton
scattering of CMB photons in the following treatment. We also ignore any
contribution to the X-ray flux density from ``upstream Compton" scattering
\citep{georganopoulos03}, whereby electrons in the jet upstream from the
hotspot inverse Compton scatter synchrotron photons produced within the
hotspot. We note that if the upstream Compton process makes a significant
contribution to the observed X-ray flux density, the SSC flux density must be
less than the observed flux density, in which case the magnetic field strength
in the hotspot would be greater than that reported in Table
\ref{table:ssc_params}, and therefore greater than the equipartition value.

\begin{deluxetable*}{ccccccccccccc}
\tabletypesize{\scriptsize}
\tablecaption{Synchrotron Self Compton Model Parameters for Northern Hotspot \label{table:ssc_params}}
\tablewidth{0pt}
\tablehead{
\multicolumn{5}{c}{Fixed Parameters} & & \multicolumn{7}{c}{Derived Parameters} \\
\cline{1-5}  \cline{7-13} \\
 \colhead{$\delta$} & \colhead{R} & \colhead{$\alpha$} & \colhead{$\nu_b$} & \colhead{$\nu_{\rm max}$} & & \colhead{B} & \colhead{B/B$_{\rm{eq}}$} & \colhead{$n_e$} &
 \colhead{$\gamma_{\rm{min}}$} & \colhead{$\gamma_{\rm{b}}$} &
 \colhead{$\gamma_{\rm{max}}$} &
 \colhead{$\alpha_{0.5\rm{keV}}^{7.0\rm{keV}}$} \\ 
 & \colhead{[pc]} & &  \colhead{[Hz]} & \colhead{[Hz]} & & [mG] & & [$\times 10^{-5}$cm$^{-3}$]& & [$\times 10^4$] & [$\times 10^5$] &  \\
}
\startdata
1.0 & 320 & 0.53 & $5 \times 10^{11}$ & $10^{14}$ & & $3.2 \pm 0.5$ & 1.5$^{+0.3}_{-0.2}$
& $2.7 \pm 0.4$ & $650 \pm 80$ & $\gtrsim \> 0.8$ &
$1.1 \pm 0.1$ & 0.5 $\pm$ 0.1  \\ 
2.0 & 320 & 0.53 & $5 \times 10^{11}$ & $10^{14}$ & & $1.0 \pm 0.2$ & $0.75 \pm 0.15$ & $2.2 \pm 0.3$ & $800 \pm 100$ & $\gtrsim  \> 1$ & $1.4 \pm 0.2$ & $0.45 \pm 0.1$ \\
3.0 & 320 & 0.53 & $5 \times 10^{11}$ & $10^{14}$ & & $0.5 \pm 0.1$ &
$0.5 \pm 0.1$ & $1.9 \pm 0.3$ & $950 \pm 100$ & $\gtrsim
\> 1.1$ & $1.6 \pm 0.2$ & $0.45 \pm 0.1$  \\ 
\enddata
\tablecomments{The quoted uncertainties on model parameters are an estimate of
  the level of uncertainty from model fitting, and correspond to the range of
  parameter values in the set of models having $\chi^2 < \chi^2_{min} +
  2.71$.} 
\end{deluxetable*}

\subsubsection{Synchrotron Self Compton Modeling} \label{sec:ssc_modeling}

To model the radio to X-ray spectral energy distribution we use the standard one-zone
SSC model: a spherical region of plasma with uniform density and magnetic
field strength. We assume that the magnetic field is ``tangled" with an
isotropic distribution of field direction. We further assume that the number
density of electrons per unit Lorentz factor is described by  

\begin{equation} \label{eqn:N_gamma}
\bar{N}(\gamma)  = \left \{ \begin{array}{ll} 0 &  \gamma < \gamma_{\rm{min}}
  , \quad \gamma > \gamma_{\rm{max}} \\ 
 \frac{K_e \gamma_b}{(a-1)} \gamma^{-(a+1)} g\left( \frac{\gamma}{\gamma_b}
 \right) & \gamma_{\rm{min}}  < \gamma < \gamma_{\rm{max}} \\ 
\end{array}
\right.
\end{equation}
where
\begin{equation} \label{eqn:g_gamma}
g \left( \frac{\gamma}{\gamma_b} \right)  = \left \{ \begin{array}{ll}  1 -
  \left(1 - \frac{\gamma}{\gamma_b} \right)^{a-1}    &  \gamma < \gamma_b \\ 
1 &  \gamma > \gamma_b \\
\end{array}
\right.
\end{equation}
$\bar{N}(\gamma) $ is the volume averaged energy distribution produced by
continuous injection of a power-law energy distribution $N(\gamma) = K_e
\gamma^{-a}$ at a shock with synchrotron cooling in a uniform magnetic field
downstream. It describes a broken power-law spectrum with the electron
spectral index smoothly changing from -a to -(a+1) at $\gamma \approx
\gamma_b$. The break in the electron spectrum at $\gamma_{b}$ corresponds to
the electron energy at which the synchrotron cooling timescale is comparable
to the dynamical timescale for electrons to escape from the hotspot. The
synchrotron cooling break is discussed further in \S
\ref{sec:break_frequency}.

For the electron energy distribution described by $\bar{N}(\gamma)$, and given
a particular radius, redshift, Doppler factor and spectral index, the
synchrotron plus self Compton spectrum is characterized by the five parameters
$K_e, B, \gamma_{\rm{min}}, \gamma_b, \gamma_{\rm{max}}$. In calculating the
model spectrum, these parameters appear in the following combinations   (see
appendix \ref{sec:appendix_ssc_equations}): 

\begin{eqnarray}
A_{\rm{syn}} &=& K_e \Omega_0^{\alpha+1} \label{eqn:A_syn} \\
A_{\rm{ssc}}  &=& K_e \gamma_b \\
\nu_1 &=& \frac{\delta}{(1+z)}\frac{3}{4 \pi}  \Omega_0 \gamma_{\rm{min}}^2 \label{eqn:nu_1} \\
\nu_b &=& \frac{\delta}{(1+z)}\frac{3}{4 \pi}  \Omega_0 \gamma_b^2  \label{eqn:nu_b}\\
\nu_2 &=& \frac{\delta}{(1+z)}\frac{3}{4 \pi}  \Omega_0 \gamma_{\rm{max}}^2 \label{eqn:epsilon_2}
\end{eqnarray}
where $\alpha = (a-1)/2$ is the radio spectral index between frequencies
$\nu_1$ and $\nu_b$, $\Omega_0$ is the non-relativistic gyro-frequency. The
parameters $\nu_1$, $\nu_b$ and $\nu_2$ correspond to the characteristic
frequency emitted by electrons with Lorentz factor $\gamma_{\rm{min}}$,
$\gamma_b$ and $\gamma_{\rm{max}}$ in a magnetic field of flux density B, and
are therefore identified with the low frequency turn-over, synchrotron cooling break and
high frequency cut-off respectively. The advantage of this formulation,
described in Appendix \ref{sec:appendix_ssc_equations}, is that it allows the
model to be specified in terms of the observed values of $\nu_1, \nu_b$ and
$\nu_2$. The parameters $A_{\rm{syn}}$ and $A_{\rm{ssc}}$ are normalization
factors for the synchrotron and SSC spectral components respectively.  

We estimate best fit values for B, $K_e$, $\gamma_{\rm{min}}$,
$\gamma_{\rm{b}}$ and $\gamma_{\rm{max}}$ using chi-squared minimization with
the following three constraints: (1) We fix the electron energy index at $a = 2.06$ so
that the spectral index $\alpha = 0.53$ for frequencies $\nu_1 << \nu <<
\nu_b$ (that is, between about 10 GHz and 100 GHz). This is the electron energy index determined from modeling the hotspot radio spectrum as described in \S
\ref{sec:whole_source_fit}. The spectral index $\alpha = 0.53$ also agrees
with the ratio of peak surface brightness (at the location of the hotspot) in
the 17.7\GHz and 20.2\GHz ATCA images. (2) We fix the radius at R = 320pc. The
radio hotspot is elongated in an approximately cylindrical shape of volume $V
\approx 4 \times 10^{57}$ m$^3$. A radius of 320pc gives a spherical model of
equal volume. (3) We fix the break frequency at $\nu_b$ = 500$\>$GHz. This is
close to the lowest break frequency allowed by the data. Higher break
frequencies are permitted but cause a worse fit to the optical data. The break
frequency is not well constrained by the data, but the results are not
sensitive to the assumed value of the break frequency. (4) We fix the upper cut-off frequency at $\nu_{\rm max} = 10^{14}$~Hz to fit the optical flux densities. 

We determined best fit parameter values while fixing the Doppler factor at
$\delta = 1, 2$ and 3. The derived model parameters are presented in
Table \ref{table:ssc_params}. The uncertainties in Table
\ref{table:ssc_params} are determined from the range of parameter values in
the set of models having $\chi^2 < \chi^2_{\rm{min}} + 2.71$. The observed
X-ray spectral index was not included in the chi-squared calculations, but the
model X-ray spectral index is consistent with the observed value within the
uncertainties.  

In Figure \ref{fig:SED} we plot the observed hotspot flux densities with the
best fit model spectra (for Doppler factors fixed at $\delta=1$ and
$\delta=3$) overlaid. The simple one-zone model with a near equipartition
magnetic field strength provides a good description of the available
data. Hardcastle et al. (2002) used more complicated spectral and spatial
models for three sources, and found that this did not have a significant
effect on the derived plasma parameters. We are therefore confident in our
parameter estimates using this ``first-order" one-zone model.

\section{Modeling the Radio Spectrum of the Entire Radio Galaxy}  \label{sec:whole_source_fit}

\begin{figure}
\plotone{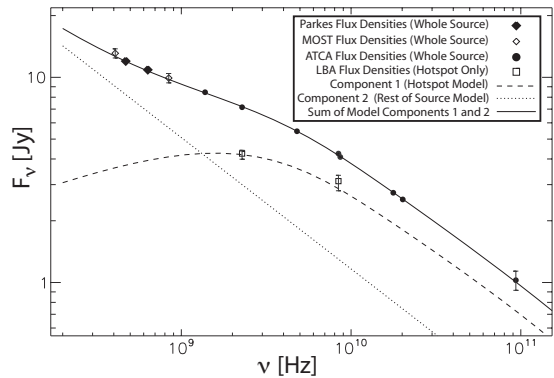}
\caption{Radio spectrum of the radio galaxy PKS~1421--490. Open diamonds
  represent flux densities of the entire radio galaxy from the Molonglo
  Synthesis Radio Telescope (MOST). Filled diamonds represent flux densities
  of the entire radio galaxy from the Parkes telescope. Filled circles
  represent flux densities of the entire radio galaxy from the ATCA. Open
  squares represent flux densities of the Northern hotspot from the LBA (see
  Table \ref{table:obs_info}). Also shown is the spectral decomposition described in
  section \ref{sec:whole_source_fit}. Component 1 (curved dashed line) is the
  angle-averaged synchrotron spectrum from a power-law electron distribution
  of the form $N(\gamma) \propto \gamma^{-2.06}$ with a low energy cut-off at
  a Lorentz factor corresponding to $\nu_1 = 2.6 \> \rm{GHz}$ (see equation
  (\ref{eqn:nu_1})). This component describes emission from the hotspot, and
  is consistent with the LBA flux densities plotted as open squares. Component
  2 (straight dotted line) is a pure power-law with spectral index 0.64. This
  component is an approximation to the emission from the rest of the
  source. The solid line is the sum of components 1 and 2. Model parameters
  are given in Table
  \ref{table:whole_source_params}.  \label{fig:whole_source_fit}}.  
\end{figure}

We now describe a consistency check for the model of the hotspot radio
spectrum in terms of a cut-off in the electron energy distribution at
$\gamma_{\rm{min}}$. This check is based on the observed flattening in the
spectrum of the entire radio galaxy (Figure \ref{fig:whole_source_fit}). In
order to test whether the observed flattening is consistent with the inferred
low energy cut-off in the electron distribution, we fit a simple two-component
model to the radio galaxy spectrum between 408\MHz and 93.5$\>$GHz. The model
components are: (1) The synchrotron spectrum produced by the electron energy distribution of equation~(\ref{eqn:N_gamma}). This component describes emission from the hotspot. (2) A pure power-law approximating emission from the rest of the source. Note that the core flux
density is negligible compared with that of the jets, lobes and hotspots, so
that no component is included to represent emission from the AGN.

The synchrotron spectrum of component 1 is calculated using equation
(\ref{eqn:j_nu_syn}). We assume the same source volume as in \S \ref{sec:ssc_modeling}, fix the lab frame break
frequency at $\nu_{\rm{b}} = 500 \> \rm{GHz}$ and the lab frame high frequency
cut-off at $\nu_{2} = 1.1 \times 10^{14} \> \rm{Hz}$, consistent with the
values determined from modeling the hotspot spectrum in \S
\ref{sec:ssc_modeling}. With these assumed values, the parameters $\nu_b$ and
$\nu_2$ do not affect the shape of the spectrum below about 100$\>$GHz, but
they weakly affect the calculation of equipartition magnetic field
strength. The spectrum of component 1 is therefore determined by the spectral
index $\alpha_1$, the turn-over frequency $\nu_1$ and the synchrotron
amplitude $A_{\rm{syn}} = K_e \Omega_0^{\alpha_1+1}$. The flux density of the
second component is of the form 
\begin{equation}
F_{\nu, \, 2} = F_{2.3\> \rm{GHz}, \, 2} \left(  \frac{\nu}{2.3 \> \rm{GHz}} \right)^{-\alpha_2}
\end{equation}
Chi-squared minimization was used to determine the best fit values for the
parameters $\alpha_1$, $A_{\rm{syn}}$, $\nu_{\rm{min}}$, $F_{2.3\> \rm{GHz},
  \, 2}$ and $\alpha_2$. The resulting model is shown in Figure
\ref{fig:whole_source_fit}. This simple two-component model provides an
excellent fit to the radio galaxy spectrum. Component 1 (describing emission
from the hotspot) is in good agreement with the LBA flux density measurements
at 2.3\GHz and 8.4$\>$GHz. We emphasize that the LBA flux density measurements
were not included in the fitting process, but are included in Figure
\ref{fig:whole_source_fit} for comparison with the spectrum of component 1.  

Again, we point out that the cut-off in number density per unit Lorentz factor below $\gamma_{\rm min}$ used in our modeling is not physically realistic, it is merely an approximation to an electron energy distribution with a sharp turn-over. The success of this simple model in simultaneously accounting for the flattening in both the hotspot spectrum and the radio galaxy spectrum supports an interpretation of the flattening in terms of a sharp turn-over in the hotspot electron energy distribution at a Lorentz factor of order $\gamma \sim 600$.

\begin{deluxetable*}{cccccccc}
\tabletypesize{\scriptsize}
\tablecaption{Model Parameters from Fitting the Radio Spectrum of the Entire
  Radio Galaxy \label{table:whole_source_params}} 
\tablewidth{0pt}
\tablehead{
\multicolumn{5}{c}{Component 1 (Hotspot)} & & \multicolumn{2}{c}{Component 2} \\
\cline{1-5} \cline{7-8} \\
\colhead{F(2.3\GHz)} & \colhead{$\alpha$} & \colhead{$\nu_1$} &
\colhead{B$_{\rm{eq}}$} & \colhead{$\gamma_{\rm{min}}^{\rm{eq}}$} & &
\colhead{F(2.3\GHz)} & \colhead{$\alpha$} \\ 
\colhead{[Jy]} & & [GHz] & [Gauss] & & & [Jy]
}
\startdata
4.15 $\pm$ 0.6 & $0.53^{+0.1}_{-0.3}$ &  $2.6^{+0.3}_{-0.5}$ &
$2.05^{+0.35}_{-0.1} \times 10^{-3}$ & $550^{+50}_{-100}$ & &3.0 $\pm$ 0.6 &
$0.65 \pm 0.1$  \\ 
\enddata
\tablecomments{The quoted uncertainties on model parameters are an estimate of
  the level of uncertainty from modelfitting, and correspond to the range of
  parameter values in the set of models having $\chi^2 < \chi^2_{min} +
  2.71$.} 
\end{deluxetable*}

\section{Synchrotron Cooling Break} \label{sec:break_frequency}

The magnetic field strength inferred from spectral modeling in section
\ref{sec:ssc_modeling} implies that there should be a synchrotron cooling
break in the hotspot radio spectrum at $\sim 1 \>$GHz if the electron energy
distribution injected at the shock is a pure power law. This is inconsistent
with the lower limit on the break frequency estimated from spectral modeling,
$\nu_b \gtrsim 500 \>$GHz. In this section we consider the production of the
cooling break, and possible reasons for the inconsistency.  

The standard continuous injection hotspot model \citep{heavens87} predicts
that the radio spectrum will steepen from $\alpha_{in}$ to $\alpha_{in} + 0.5$
at a frequency, $\nu_b$, corresponding to the electron energy at which the
synchrotron cooling time-scale $\tau_{\rm{cool}}$ is equal to the dynamical
time-scale $\tau_{\rm{esc}}$ for electrons to escape the hotspot. The break
frequency is an important constraint on the physics of the hotspot. In
general, it depends on the magnetic field strength, hotspot radius, outflow
velocity, Doppler factor and the presence or absence of a re-acceleration
mechanism within the hotspot.  We consider a model in which the escape
time-scale $\tau_{esc}$ is the time taken for the flow to cross the hotspot
and the cooling time-scale is the synchrotron half-life. Let R be the hotspot
radius, $v=\beta_{\rm{f}} c$ the flow velocity within the hotspot (note that
this is not the same as the advance velocity of the hotspot),
$\delta_{\rm{f}}$ the corresponding Doppler factor of the flow within the
hotspot, and $U_B$ the magnetic field energy density ($U_B = B^2 / 2 \mu_0$ in
S.I. units, $U_B = B^2 / 8 \pi$ in c.g.s. units).  
\begin{eqnarray}
\tau_{\rm{esc}} &=& \frac{2 R}{\beta_{\rm{f}} c} \\
\tau_{\rm{cool}} &=& \frac{\gamma}{| d \gamma/ dt |} \\
&=& \frac{3 m_e c}{4 \sigma_T U_B\, \gamma}
\end{eqnarray}
Equating the two time-scales and combining with equation (\ref{eqn:nu_b}) for
the break frequency in terms of the break Lorentz factor, we obtain the
following expression for the break frequency 
\begin{eqnarray} 
\nu_{\rm{b}} &\approx& 60 \, \frac{\delta_{\rm{f}}^{\frac{7 + 4 \alpha}{\alpha
      + 1}} \, \beta_{\rm{f}}^2  }{(1+z)}    \left(
\frac{B_{\delta_{\rm{f}}=1}}{\rm{mG}}  \right)^{-3}   \left(
\frac{R}{\rm{kpc}} \right)^{-2} \rm{GHz} \\ 
&=& 60 \, \frac{\delta_{\rm{f}}^6 \, \beta_{\rm{f}}^2}{(1+z)}    \left(
\frac{B_{\delta_{\rm{f}}=1}}{\rm{mG}}  \right)^{-3}   \left(
\frac{R}{\rm{kpc}} \right)^{-2} \rm{GHz}  \qquad (\alpha =
0.5) \label{eqn:nu_b_doppler_factor} 
\end{eqnarray}
where $B_{\delta_{\rm{f}}=1}$ is the magnetic field strength derived from SSC
modeling under the assumption $\delta_{\rm{f}} = 1$. For a Doppler factor
$\delta_{\rm{f}} = [ \Gamma (1 - \beta_{\rm{f}} \cos \theta)]^{-1}$, the
magnetic field strength estimated from SSC modeling is reduced by a factor of
approximately $\delta_{\rm{f}}^{-\frac{\alpha+2}{\alpha+1}}$
\citep[eg.][]{worrall06}. Equation (\ref{eqn:nu_b_doppler_factor}) exhibits a
strong dependence on the Doppler factor because of the strong dependence of
the break frequency on the magnetic field strength.   

Let us first consider the production of the cooling break in a hotspot
associated with a strong relativistic normal shock in which the post-shock
velocity $\beta_{\rm{f}} \approx 0.3$, $\delta_{\rm{f}} \approx 1$, redshift
$z=0.663$, magnetic field $B = 3 \> \rm{mG}$ (as determined from SSC modeling
in \S \ref{sec:ssc_modeling}) and radius R=0.3kpc (half the geometric mean of
the longest and shortest angular sizes of the 2.3\GHz LBA image). For such a
model, the predicted break frequency is $\nu_b \approx 1 \>$GHz. This is
inconsistent with the lower limit from spectral modeling, $\nu_b \gtrsim 500
\>$GHz. Moreover, the break frequency estimated from spectral modeling is
inconsistent with the proposed correlation between break frequency and
equipartition magnetic field strength \citep[eg.][]{brunetti03, cheung05}
which also predicts a break frequency $\nu_b \sim 1 \> \rm{GHz}$.  The
discrepancy between predicted break frequency and the observed lower limit
would be alleviated if the magnetic field strength were $\lesssim 0.15$ of the
value estimated from SSC modeling. However, if this were the case, the model
SSC spectrum would over-predict the observed X-ray flux density.  

Let us now consider the effect of Doppler beaming on the observed break
frequency as a possible means of resolving this difficulty. Assuming a
post-shock flow velocity $\beta_{\rm{f}}^2 \sim (1 - \delta_{\rm{f}}^{-2})$
and a spectral index of $\alpha \sim 0.5$, a moderate Doppler factor
$\delta_{\rm{f}} \gtrsim 1.9$ is sufficient to increase the predicted break
frequency above 500$\>$GHz, while maintaining agreement between the SSC model
spectrum and the observed flux densities. 

There is also the possibility that distributed re-acceleration within the
hotspot is affecting the production of the cooling break.
\citet{meisenheimer97}  have suggested that distributed re-acceleration is
required to explain the spectra of the so-called low-loss hotspots. These are
hotspots whose spectra are characterized by a power-law with $\alpha \approx
0.6 - 0.8$ that extends to high frequency ($\nu > 10^{12} \> \rm{Hz}$) without
the predicted break in the spectrum. Distributed re-acceleration has also been
proposed to explain the diffuse infrared/optical emission observed around some
hotspots \citep{prieto02, roser87, meisenheimer03}, as well as the variation
in the X-ray spectral index around the hotspots of Cygnus A
\citep{balucinska-church05}, and the existence of flat radio spectrum regions
distributed over much of the hotspot area in Cygnus A \citep{carilli99}. The
favoured mechanism for re-acceleration is stochastic (second order Fermi)
acceleration via magnetohydrodynamic turbulence \citep{meisenheimer97,
  prieto02, balucinska-church05}.  

Lastly, it is possible that there is more than one site of particle injection,
or that the hotspot is not in a steady state.

\section{Is Doppler Beaming Significant in the Northern Hotspot of PKS~1421--490?} \label{sec:Doppler}

The aim of this section is to assess the likelihood that emission from the the northern hotspot of PKS1421-490 is Doppler beamed. To do so, in \S \ref{sec:arguments_for} we lay out the evidence for Doppler beaming in the northern hotspot, then discuss results of numerical simulations and radio studies that indicate Doppler beaming may be significant in hotspots of radio galaxies and quasars. In \S \ref{sec:inclination_angle} we consider the angle to the line of sight of PKS1421-490, which should be small if Doppler beaming is to be important. In \S \ref{sec:doppler_factor} we estimate the magnitude of the Doppler factor that would be required to account for various properties of the hotspot. In \S \ref{sec:arguments_against} we discuss two possible arguments against Doppler beaming. 

\subsection{Arguments for Doppler Beaming} \label{sec:arguments_for}

The northern hotspot of PKS~1421--490 is extremely luminous at both radio and
X-ray wavelengths. The X-ray luminosity between 2 and 10 keV, $L_{2-10\>
  \rm{keV}} = 3 \times 10^{44} \> \rm{ergs \> s^{-1}}$, is comparable to the
X-ray luminosity of the entire jet of PKS~0637--752, without relativistic
corrections. The peak radio surface brightness is hundreds of times
greater than that of the brightest hotspot in Cygnus A
\citep{carilli99}. Consequently, the equipartition magnetic field strength for a Doppler factor of unity is
greater by a factor of $\sim$5 - 10, and the minimum energy density is greater
by a factor of $\sim$50 than values typically evaluated for bright hotspots in
other radio galaxies \citep{meisenheimer97, tavecchio05, kataoka05}. The
northern hotspot contributes $60 \%$ of the total source flux density at
2.3$\>$GHz, and 75$\%$ at 8$\>$GHz. Identifying the peak of region C in the
ATCA image as the counter hotspot, we estimate the
hotspot to counter hotspot flux density ratio to be $\rm{R_{hs}} \sim 300$ at
20$\>$GHz. In the {\it Chandra} X-ray band, the counter-hotspot is undetected,
and we conservatively estimate $\rm{R_{hs}} > 100$ at X-ray
wavelengths. These are all indications that the hotspot emission may be
Doppler beamed. Moreover, we have shown in \S \ref{sec:break_frequency} that Doppler beaming may account for the absence of a synchrotron cooling break below 500~GHz. We now discuss the results of numerical simulations and radio studies that indicate Doppler beaming may be important in hotspots of radio galaxies.  

Numerical simulations of supersonic jets in 2 and 3 dimensions
\citep[eg.][]{aloy99, norman96, komissarov96, tregillis01} suggest that flow
speeds in and around hotspots can be much larger than those expected from the
1D strong shock model, because the shock structure at the jet termination is
more complex than a single terminal Mach disk. The simulated jets undergo
violent structural and velocity changes near the jet head due to pressure
variations in the turbulent cocoon. These violent changes in the jet affect
the hotspot structure, and may result in an oblique shock (or shocks) near the hotspot \citep{aloy99, norman96}. The post-shock velocity of an oblique shock can be
much higher than the post-shock velocity of a normal shock if the angle
between the flow velocity and the shock normal is close to the Mach
angle. Therefore, the instantaneous flow velocity through the hotspot may be high enough to produce significant Doppler effects \citep{aloy99}. 

If the terminal shock is not highly oblique, the post-shock velocity may be relativistic if the jet contains a dynamically important magnetic field. The magnetic field can reduce the shock compression ratio and result in a higher post-shock Lorentz factor than that in an un-magnetized shock \citep[see eg.][]{double04}. The post-shock velocity in a magnetized shock depends on the the angle between the magnetic field and the shock plane, the equation of state in the pre- and post-shock plasma, and the magnetization parameter $\sigma = B^2/4 \pi \rho c^2$ \citep{double04}. In the case where the magnetic field is perpendicular to the jet direction, significant post-shock Lorentz factors ($\Gamma_2 \gtrsim 2$) can be achieved if $\sigma \gtrsim 3$, depending on the equation of state. We suggest that, given the high magnetic field strength in the northern hotspot of PKS~1421--490, magnetic cushioning of the terminal shock due to the presence of a strong magnetic field in the jet may be important. 

In addition to the results of numerical simulations, observational evidence
also indicates that Doppler beaming of hotspot emission may be significant. For
example, the brighter and more compact hotspot is generally found on the side
of the source with the brighter kpc-scale jet  \citep[eg.][]{bridle94,
  hardcastle98}. This effect is more evident in samples of quasars than in
samples including low power sources, which suggests that the observed
correlation between hotspot brightness and jet brightness is related to
Doppler beaming \citep{hardcastle98}. However, \citet{hardcastle03} suggest
that only moderate hotspot flow velocities ($\beta \sim 0.3$) are required to
account for this observed correlation. \citet{dennett-thorpe97} found that
regions of high surface brightness in the lobes of radio galaxies have flatter
radio spectra on the side corresponding to the brighter jet. They suggest that
Doppler shifting of a curved hotspot spectrum may produce such a
correlation. Again, only moderate flow speeds of $\beta \lesssim 0.5$ are
required to account for this correlation
\citep{ishwara-chandra00}. \citet{georganopoulos03} have suggested that
deceleration of a relativistic flow from $\Gamma \sim 3$ to $\Gamma \sim 1$ in
hotspots can explain the wide range of observed hotspot SEDs as being purely an effect
of source inclination. However, \citet{hardcastle03} and \citet{hardcastle04}
have contested this interpretation. Rather, they argue, the shape of the hotspot SED depends only on the hotspot radio luminosity. 

\subsection{Jet Inclination Angle} \label{sec:inclination_angle}

We now consider the angle to the line of sight for PKS~1421--490, if Doppler
beaming is to be important.  

The active galactic nucleus of PKS~1421--490 exhibits broad emission lines (see \S \ref{sec:region_B_spectrum}). On
the basis of the unified scheme for active galaxies, we therefore expect the
angle to the line of sight to be less than $\sim$45$^{\circ}$
\citep{urry95}. Another indication of a small angle to the line of sight is
the existence of a $60^{\circ}$ bend in the northern jet, approximately 5
arcseconds (35kpc) from the AGN at the western end of the ridge of emission
extending west from the hotspot in Figure \ref{fig:ATCA}. Such a large jet
deflection is hard to understand if it is indicative of the true bending
angle. The well known resolution to this problem is that the jet is viewed
close to the line of sight, and the effect of projection causes a relatively
small jet deflection to appear much larger than it actually is.  

\subsection{Estimates of the Doppler Factor} \label{sec:doppler_factor}

We now consider the magnitude of the Doppler factor that would be required to
account for the various observed properties.  

Let $\rm{R_{hs}}$ be the hotspot to counter hotspot flux density ratio,
$\beta_{hs}$ the bulk flow velocity in the hotspot divided by the speed of
light, $\theta$ the jet angle to the line of sight, and $\alpha$ the spectral
index. If the two hotspots of PKS~1421--490 are identical, and the difference
in flux density is purely the result of relativistic beaming, then  

\begin{equation}   \label{eqn:counter-HS_ratio}
\beta_{hs} \cos \theta = \frac{\rm{R_{hs}}^{\frac{1}{2 + \alpha}} -
  1}{\rm{R_{hs}}^{\frac{1}{2 + \alpha}} + 1} 
\end{equation}
\citep[eg.][pg. 73]{deyoung02}. 

The observed hotspot to counter-hotspot flux density ratio is $\rm{R_{hs}}
\sim 300$ at 20.2$\>$GHz, hence: $\beta_{\rm hs} \cos
\theta \sim 0.81$, $\Gamma_{\rm hs} > 1.7$, and $\theta < 36^{\circ}$. A moderate
Lorentz factor of $\Gamma_{\rm hs} \gtrsim 1.7$ can account for the observed
hotspot flux density ratio. The bend in
the northern jet means that we cannot assume the same inclination angle for
the jet and counter-jet, so equation (\ref{eqn:counter-HS_ratio}) does not
strictly apply, but the above calculations serve to illustrate that the
required Lorentz factor is not large. If the jet is angled close to the line of sight, the real difference in inclination angle between jet and counter-jet may not be large. 

It should be noted that there is a
difference between the times at which we see the two hotspots. In the case of
PKS~1421--490 this difference is approximately $(3 \times 10^5 / \tan \theta)
\>$yrs, where $\theta$ is the angle to the line of sight. In equation
(\ref{eqn:counter-HS_ratio}) there is an implicit assumption that the brightness
of the hotspots remain constant over a time-scale of approximately $10^5 -
10^6$ years.

We summarize below the estimates of the Doppler factor required to account for
various properties of the hotspot. 

\begin{itemize}
\item[1.] A bulk Lorentz factor $\Gamma \gtrsim 1.7$
  is required to account for the observed hotspot to counter hotspot flux
  density ratio of $\rm{R_{hs}} \sim 300$. If the bend in the Northern jet is
  such that a decrease in inclination angle is produced, the required bulk
  Lorentz factor is lower.  
\item[2.] A Doppler factor $\delta \gtrsim 1.9$ is required to account for the
  observed lower limit on the break frequency $\nu_{\rm{b}} \gtrsim 500 \>
  \rm{GHz}$ (see \S \ref{sec:break_frequency}). 
\item[3.] A Doppler factor $\delta \sim 3$ is required to reduce the SSC model
  magnetic field strength to a value comparable to that calculated for other
  radio galaxies ($\sim 100 - 500 \> \rm{\mu G}$) \citep{kataoka05}. However,
  such agreement is not essential since some variation in the radio galaxy
  population would be expected.   
\end{itemize}

\subsection{Arguments Against Doppler Beaming} \label{sec:arguments_against}

In \S \ref{sec:region_HS2} we argued that the broad emission to the west of the hotspot peak (region HS2) may be associated with turbulent back-flow in the cocoon, and therefore cannot be Doppler beamed. As further discussed in \S \ref{sec:region_HS2}, if the high surface brightness of region HS2 is not the result of Doppler beaming, then it seems unreasonable to argue that the high surface brightness of region HS1 is the result of Doppler beaming.  

\begin{figure}
\epsscale{0.75}
\begin{center}
\plotone{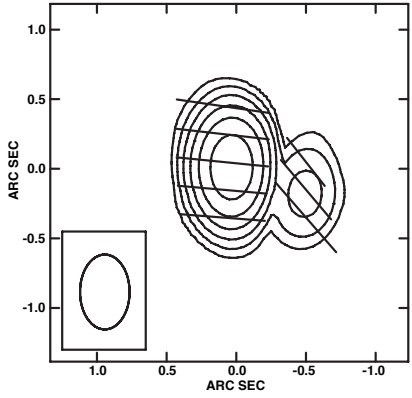}
\caption{20.16GHz ATCA contours of polarized intensity with polarization
  E-vectors overlaid. The length of the polarization E-vectors is proportional
  to the fractional polarization, with 1 arcsecond corresponding to 15$\%$
  fractional polarization. The polarization vectors have been rotated by -1.8
  degrees to account for the observed rotation measure of $\sim 140$
  rad/m$^2$. The main peak corresponds to the hotspot (regions HS1 and HS2),
  while the secondary peak is associated with region J1 in Figure
  \ref{fig:LBA}, presumably associated with a jet knot. Both regions are
  approximately $8\% - 10\%$ polarized. The map peak is 0.19 Jy/beam.  Contour
  levels: 3.5$\>$mJy/beam $\times$ (1, 2, 4, 8, 16, 32, 64).  Beam FWHM: 0.54
  $\times$ 0.36 arcsec. The scale of this image is 7.0
  kpc/arcsecond. \label{fig:1421_pcntr}}.  
\end{center}
\end{figure}

Another possible argument against Doppler beaming comes from interpreting the radio polarization. Figure \ref{fig:1421_pcntr} is a contour map of the linearly polarized intensity at 20$\>$GHz in the vicinity of the hotspot, with polarization position angle indicated by the vectors overlaid. The main peak in the
contour map is associated with the hotspot (regions HS1 and HS2). The offset
of the secondary peak relative to the main peak places it at the same position
as region J1 in Figure \ref{fig:LBA}. The position angle of the E-vectors in
the hotspot indicate that the magnetic field (perpendicular to the E-vectors)
is aligned nearly perpendicular to the jet direction. The magnetic field
direction is often identified with the shock plane, because the component
of the magnetic field in the plane of the shock is amplified, while the
component of magnetic field perpendicular to the shock plane is conserved. Figure \ref{fig:1421_pcntr} therefore indicates that the terminal shock is not highly oblique. If the terminal shock is not highly oblique the post-shock velocity cannot be highly relativistic unless the magnetic field is dynamically important (see \S \ref{sec:arguments_for} and \citet{double04}). If the magnetic field is dynamically important, this argument against Doppler beaming based on the polarization position angle is not valid.

\section{Interpreting the Low Frequency Flattening in the Radio Spectrum of the Northern Hotspot} \label{sec:gamma_min} 

The aim of this section is to consider the implications of the observed flattening in the hotspot radio spectrum discussed in \S \ref{sec:low_frequency_flattening} and illustrated in Figure \ref{fig:plot_showing_flattening}. We consider two possible
mechanisms for producing a flattening in the electron energy
distribution at $\gamma_{\rm min} \sim 650$. The first mechanism we consider
is the dissipation of jet bulk kinetic energy. Dissipation of the jet kinetic energy depletes low
energy particles and produces a turn-over in the electron spectrum at a characteristic energy that depends on a number of parameters, including the jet Lorentz factor. The second mechanism we consider is a transition between two distinct acceleration mechanisms. 

We note that the inferred value of $\gamma_{\rm{min}}$ is only weakly dependent on the
assumed Doppler factor $\delta$ because Doppler beaming affects the
calculation of the magnetic field strength ($B \propto
\delta^{-(2+\alpha)/(1+\alpha)}$) as well as the rest frame emission frequency
corresponding to $\gamma_{\rm{min}}$ ($\nu^{\prime} \propto \delta^{-1}
\nu$). The value of $\gamma_{\rm{min}}$ is therefore approximately
proportional to $\delta^{1/3}$ (see Table \ref{table:ssc_params}).

\subsection{Dissipation of Jet Energy}

As an illustrative calculation, we consider the dissipation of jet energy in a cold, un-magnetised proton/electron jet. The analysis is effectively done in two steps: First, we use the conservation of energy and particles to calculate the mean Lorentz factor in the hotspot as a function of jet Lorentz factor. We then relate the mean electron Lorentz factor to the peak Lorentz factor by assuming a particular form for the electron energy distribution. We do not specify the process by which the electrons and protons equilibrate. However, recent particle-in-cell simulations demonstrate that protons and electrons equilibrate in un-magnetised collisionless shocks \citep{spitkovsky08}. 

The aim of this calculation is to estimate the jet Lorentz factor required to produce a turn-over in the electron energy distribution at $\gamma_{\rm min} \sim 650$ if the jet bulk kinetic energy is carried by protons and efficiently transfered to electrons in the hotspot. This analysis can easily be extended to include different jet compositions, different proton to electron energy density ratios and the effects of the magnetic field. 

\subsubsection{Model Assumptions and Definitions} \label{sec:model_assumptions}

The relevant quantities are defined as follows: $\gamma $ is the Lorentz factor of an individual particle measured in the plasma rest frame, $\gamma_p$ is the electron Lorentz factor at the peak of the electron energy distribution, $\Gamma$ is the bulk Lorentz factor of the plasma, $\beta = \sqrt{1 - \Gamma^{-2}}$ is the corresponding plasma speed in units of the speed of light c, $\theta$ is the angle between the plasma velocity and the line of sight, $\delta = [\Gamma (1 - \beta \cos \theta)]^{-1}$ is the Doppler factor, $N(\gamma)$ is the number density of electrons per unit Lorentz factor, $\langle \gamma \rangle = \int \gamma N(\gamma) d \gamma / \int N(\gamma) d \gamma$ is the mean Lorentz factor, $n = \int N(\gamma) d\gamma$ is the number density, $\epsilon = \left( \langle \gamma \rangle - 1 \right) n m c^2$ is the internal energy density, $\rho = n m$ is the rest mass density in the plasma rest frame, $p$ is the pressure and $w$ is the relativistic enthalpy density. The relativistic enthalpy density is
\begin{equation}
w = \epsilon + p + \rho c^2
\end{equation}

We assume the plasma is comprised of electrons (subscript e) and protons (subscript p). Quantities with the subscript 1 refer to the jet plasma, while quantities with a subscript 2 refer to the hotspot plasma. 

We assume that the relativistic enthalpy density in the jet is dominated by the rest mass energy density of the the proton component, so that $w_{1} \approx \rho_{1, p} c^2$. This assumption is valid provided $\langle \gamma \rangle_{1, e} << m_p/m_e$. 
 
We assume that the electron population is ultra-relativistic ($\epsilon_{e} = 3 p_{e}$), and that the proton population in the hotspot plasma is, at best, only mildly relativistic, and can be approximated as a thermal gas ($p_{2, p} = \frac{2}{3} \epsilon_{2, p}$). We further assume that the protons and electrons equilibrate so that $\epsilon_{2, p} = \epsilon_{2, e}$. Hence, the hotspot pressure and enthalpy are $p_2 = \epsilon_{2, e}$ and $w_2 = \rho_2 c^2 + 3 \epsilon_{2, e}$. 

\subsubsection{Conservation Equations}

The equations for the conservation of energy and particles are, respectively
\begin{eqnarray}
 A_1 \Gamma_1^2 \beta_1  w_1   &=& A_2 \Gamma_2^2 \beta_2  w_2  \label{eqn:F_E}  \\
 A_1 \Gamma_1 \beta_1 n_1  &=& A_2 \Gamma_2 \beta_2 n_2 \label{eqn:F_p} 
\end{eqnarray}
where A is the jet cross-sectional area. Dividing equation (\ref{eqn:F_E}) by equation (\ref{eqn:F_p}), we can write:
\begin{equation} \label{eqn:F_E_on_F_p}
\Gamma_1 \frac{w_1}{\rho_1 c^2}  = \Gamma_2 \frac{w_2}{\rho_2 c^2}
\end{equation}

\subsubsection{Peak Lorentz Factor}

Combining equation (\ref{eqn:F_E_on_F_p}) with our model assumptions described above we find
\begin{eqnarray}
\Gamma_1 &\approx& \Gamma_2 \left( 1 + \frac{3 \epsilon_{2, e}}{\rho_2 c^2}  \right) \\
&\approx&  \Gamma_2 \left( 1 + 3 \frac{m_e}{m_p}   \langle \gamma \rangle  \right)
\end{eqnarray}
where we have made the substitution $\rho_2 \approx n_p m_p$ and $\epsilon_{2, e} = \langle \gamma \rangle n_e m_e c^2$.

Let us introduce the parameter $\chi \equiv \langle \gamma \rangle/\gamma_p$ which is the ratio of the mean electron Lorentz factor in the hotspot ($\langle \gamma \rangle$) to the electron Lorentz factor at the peak of the electron energy distribution ($\gamma_p$). Then
\begin{equation}
\gamma_p = \frac{1}{3} \frac{m_p}{m_e} \left( \frac{\Gamma_1}{\Gamma_2} - 1 \right) \chi^{-1}
\end{equation}

\begin{figure}
\begin{center}
\plotone{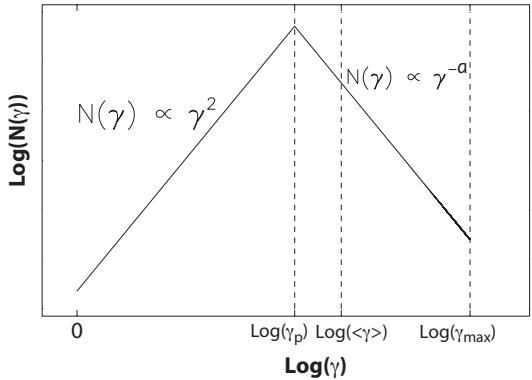}
\caption{Plot of the assumed form of the electron energy distribution used to
  relate the mean of the distribution $\langle \gamma \rangle$ to the mode
  $\gamma_{\rm{p}}$ (see equations (\ref{eqn:assumed_n_gamma}) and
  (\ref{eqn:approx_n_gamma})). In \S \ref{sec:modelling}, when modeling the
  hotspot spectrum, we approximated the electron energy distribution with a
  broken power law between $\gamma_{\rm{min}}$ and $\gamma_{\rm{max}}$, and
  number density set to zero outside this range. Here we assume that initially
  the (injected) electron spectrum is a single (un-broken) power-law between
  $\gamma_{\rm{p}}$ and $\gamma_{\rm{max}}$, with $N(\gamma) \propto \gamma^{2}$ below $\gamma_{\rm p}$. The instantaneous cut-off at
  $\gamma_{\rm{min}}$ used for spectral modeling is an approximation to the turnover in the electron
  energy distribution. We identify $\gamma_{\min}$ with
  the peak of the electron energy distribution, which here we define as
  $\gamma_p$.  \label{fig:assumed_electron_distribution}}.  
\end{center}
\end{figure}

In order to estimate the parameter $\chi$ we assume that the electron distribution below $\gamma_p$ can be approximated by the low energy tail of a relativistic Maxwellian, and above $\gamma_p$ the electron distribution (before cooling via synchrotron emission) is a power-law extending from the peak to a maximum Lorentz factor $\gamma_{\rm{max}}$ (see Figure \ref{fig:assumed_electron_distribution}).

\begin{equation} \label{eqn:assumed_n_gamma}
N(\gamma)  = \left \{ \begin{array}{ll} K_e \gamma_{\rm{p}}^{-(a+2)} \gamma^2 & 1  < \gamma < \gamma_{\rm{p}} \\
K_e \gamma^{-a} & \gamma_{\rm{p}} < \gamma < \gamma_{\rm{max}} \\
0 &  \quad \gamma > \gamma_{\rm{max}} \\
\end{array}
\right.
\end{equation}
Using this particular form for the energy distribution, the value of $\chi$ is a function of the three parameters a, $\gamma_p$ and $\gamma_{\rm max}$. If $a=2$, the ratio of mean Lorentz factor to the peak Lorentz factor reduces to the following simple algebraic form
\begin{equation} \label{eqn:approx_n_gamma}
\chi \approx \frac{3}{16}  + \frac{3}{4} \rm{ln}\left( \frac{\gamma_{\rm{max}}}{\gamma_{\rm{p}}} \right) 
\end{equation}
provided $\gamma_{\rm{p}} >> 1$. From our analysis of the spectrum of the northern hotspot of PKS~1421--490 in \S \ref{sec:ssc_modeling} we have $\gamma_p \approx 650$, $\gamma_{\rm{max}} \approx 1.2 \times 10^5$ and $a=2.06$ so that $\chi  \approx 4$. Therefore in order to produce a turn-over in the electron energy distribution at $\gamma_{\rm min} \sim 650$ the jet must have a bulk Lorentz factor $\Gamma_1 \gtrsim 5$. This value of jet Lorentz factor is consistent with jet Lorentz factors inferred from modelling the radio to X-ray spectra of quasar jets on kpc-scales \citep{tavecchio00, schwartz06, kataoka05}. If $\Gamma_2 > 1$, or the jet contains some fraction of positron/electron pairs, or the electrons do not reach equilibrium with the protons, or we consider the effect of the magnetic field, then the energy requirements increase, and so too must $\Gamma_1$.

It was noted in \S \ref{sec:intro} that while only a small number of hotspots have provided direct estimates of $\gamma_{\rm{min}}$, they are all distributed around a value of $\gamma_{\rm{min}} \approx 600$ to within a factor of 2 (excluding the value $\gamma_{\rm{min}} \sim 10^4$ indirectly estimated by \citet{blundell06}). The value of the parameter $\chi$ is weakly dependent on the electron spectrum, and in general should be within the range $\chi \sim 2 - 6$. Therefore, dissipation of bulk kinetic energy associated with relativistic proton/electron jets with bulk Lorentz factors of order $\Gamma_1 \gtrsim 5$ can provide a natural explanation for the inferred turn-overs in electron spectra at $\gamma_{\rm{min}} \sim 300 - 1000$. 

Our analysis indicates that the value of $\gamma_{\rm{min}} \sim 10^4$ inferred by \citet{blundell06} for the hotspot of the radio galaxy 6C0905+3955 would require a jet Lorentz factor $\Gamma_{\rm jet} \gtrsim 16 \chi + 1$. If $\chi \gtrsim 2$ (note that $\alpha \sim 0.7$ for this hotspot, and the synchrotron spectrum extends from radio through to soft X-ray frequencies  \citep{erlund08}), then the required jet Lorentz factor is $\Gamma_{\rm jet} \gtrsim 35$.

\subsubsection{Electron/Positron Jet}

Let us now consider the case of a pure electron/positron jet. In this case, assuming an ultra-relativistic equation of state in the jet and hotspot, the ratio of equations (\ref{eqn:F_E}) and (\ref{eqn:F_p}) implies
\begin{equation}
\Gamma_1 \gtrsim \frac{\langle \gamma \rangle_{2, e}}{\langle \gamma \rangle_{1, e}}
\end{equation}
\citet{uchiyama05} estimate a mean Lorentz factor of $\langle \gamma \rangle_{1, e} \approx 50$ in the jet of PKS~0637--752. If a similar mean Lorentz factor applies to the jet of PKS~1421--490 then the required jet bulk Lorentz factor is $\Gamma_1 \gtrsim 50$. 

\subsection{Pre-Acceleration: Cyclotron Resonant Absorption?}

We now consider an alternative explanation for the flattening of the electron
energy distribution. The observed change in slope may be the result of a
transition between two different acceleration mechanisms: a pre-acceleration
process producing a relatively flat electron spectrum at low energy, and
diffusive shock acceleration acting at higher energy producing an electron distribution $N(\gamma) \propto \gamma^{-a}$ with $a \sim 2$. We have not modeled the spectrum in terms of such a scenario, but this model cannot yet be ruled out.  

One interesting candidate for the pre-acceleration mechanism is that
described by \citet{hoshino92} and \citet{amato06}. They have shown that in a
relativistic, magnetized, collisionless shock with an electron-positron-proton
plasma there can be efficient transfer of energy from protons to
leptons via the resonant emission and absorption of
electromagnetic waves at high harmonics of the proton cyclotron
frequency. This process produces a particle distribution described by a
relativistic Maxwell distribution at low energies ($\gamma < \Gamma_{\rm jet}$)
and a relatively flat ($a < 2$) power-law component extending from $\sim
\Gamma_{\rm jet}$ to $\sim \Gamma_{\rm jet}(m_p / m_e)$. The electron energy index of the
power-law component is sensitive to the plasma composition. The theoretical
maximum Lorentz factor attained via this acceleration mechanism ($\sim
\Gamma_{\rm jet} (m_p / m_e)$) is set by resonance with the fundamental proton
cyclotron frequency. However, the upper cut-off energy determined from the
results of particle-in-cell simulations is somewhat lower than the theoretical maximum
\citep{amato06}. Therefore, the observed flattening in the hotspot radio
spectrum may be associated with a transition between the cyclotron resonant
absorption mechanism and diffusive shock acceleration.  \citet{stawarz07} have
suggested this interpretation for the hotspot spectra in Cygnus A.  

\section{Conclusions} \label{sec:conclusions}

Long Baseline Array (LBA) imaging of the z=0.663 broad
line radio galaxy PKS~1421--490 has revealed a compact (400 pc diameter), high
surface brightness hotspot at a projected distance of approximately 40 kpc
from the active galactic nucleus. The isotropic X-ray luminosity of the hotspot, $L_{2-10\> \rm{keV}} =
3 \times 10^{44} \> \rm{ergs \> s^{-1}}$, is comparable to the isotropic X-ray
luminosity of the entire X-ray jet of PKS~0637--752, and the peak radio surface brightness is hundreds of times greater than that of the brightest hotspot in Cygnus A. We successfully modeled the radio to X-ray spectral energy distribution using a standard one-zone synchrotron self Compton model with a
near equipartition magnetic field strength of 3 mG. There is a strong brightness asymmetry between the approaching and receding hotspots, and
the hot spot spectrum remains flat ($\alpha \approx 0.5$) well beyond the predicted
cooling break for a 3~mG magnetic field,
indicating that the hotspot emission may be Doppler beamed. We suggest that a high plasma velocity beyond the terminal jet shock could be the result of a dynamically important magnetic field in the jet, resulting in Doppler boosted hotspot emission. However, some aspects of the hotspot morphology may argue against an interpretation involving significant Doppler beaming. LBA observations at 1.4~GHz will be required to further investigate the hotspot morphology. 

There is a change in the slope of the hotspot radio spectrum at GHz
frequencies. We successfully modeled this feature by incorporating a cut-off in
the electron energy distribution  at $\gamma_{\rm{min}} \sim 650$ (assuming a Doppler factor of unity). If the hotspot emission is Doppler beamed with Doppler factor $\delta$, the low energy cut-off is $\gamma_{\rm min} \approx 650 \; \delta^{1/3}$. We have made use of the equations for the conservation of energy and particles in an un-magnetised proton/electron jet to obtain a general expression that relates the peak in the hotspot electron energy distribution to the jet bulk Lorentz factor. We have shown that a sharp decrease in electron number density below a Lorentz factor of about 650 would arise from the dissipation of bulk kinetic energy in an electron/proton jet with bulk Lorentz
factor $\Gamma_{\rm jet} \gtrsim 5$. This value of jet Lorentz factor is consistent with jet Lorentz factors inferred from modelling the radio to X-ray spectra of quasar jets on kpc-scales \citep{tavecchio00, schwartz06, kataoka05}. These results are of particular interest given that
similar values of $\gamma_{\rm{min}}$ have been estimated for several other
hotspots. Our analysis indicates that the value of $\gamma_{\rm{min}} \sim 10^4$ inferred by \citet{blundell06} for the hotspot of the radio galaxy 6C0905+3955 would require a jet Lorentz factor $\Gamma_{\rm jet} \gtrsim 35$. 

An alternative explanation for the low frequency flattening in the radio spectrum of the northern hotspot
of PKS~1421--490 may be that it is associated with the transition between a
pre-acceleration mechanism, such as the cyclotron resonant process described
by \citet{hoshino92} and \citet{amato06}, and diffusive shock acceleration.  

Future LBA observations at 1.4GHz will help to constrain the low energy end of
the electron energy distribution,  and infra-red observations are required to
constrain the high frequency end of the synchrotron spectrum. More sophisticated models of the electron energy distribution will be required in future studies, to
test the hypothesis that the flattening in the radio spectrum is associated
with a transition between two distinct acceleration mechanisms.

\acknowledgments

L.E.H.G would like to thank John Kirk for very helpful discussions on the
topic of particle acceleration, and Emil Lenc for supplying the \emph{cordump}
patch for DIFMAP. L.E.H.G would also like to thank the Grote Reber Foundation
for financial support. The Australia Telescope Compact Array and Long Baseline
Array are part of the Australia Telescope which is funded by the Commonwealth
of Australia for operation as a National Facility managed by CSIRO. This paper
includes data gathered with the 6.5 meter Magellan Telescopes located at Las
Campanas Observatory, Chile. This publication makes use of data products from the Two Micron All Sky Survey, which is a joint project of the University of Massachusetts and the Infrared Processing and Analysis Center/California Institute of Technology, funded by the National Aeronautics and Space Administration and the National Science Foundation.

{\it Facilities:} \facility{LBA},\facility{ATCA}, \facility{Magellan:Baade (IMACS)}.

\appendix

\section{Equations for Synchrotron and SSC model Flux Density} \label{sec:appendix_ssc_equations}

\subsection{Angle Averaged Synchrotron Flux Density for an Arbitrary Electron Energy Distribution}

The standard expressions for the synchrotron spectrum produced by an arbitrary
electron energy distribution contain a dependence on the angle between the
line of sight and the magnetic field \citep[see eg.][]{worrall06}. If the
magnetic field direction changes significantly throughout the volume in which
the observed flux is produced, it is appropriate to use the angle averaged
emission spectrum to model the source. In this appendix we present a formal
way of calculating the angle averaged synchrotron spectrum for an arbitrary
electron energy distribution. We then give the expression used in calculating
the flux density from the volume averaged shock distribution described in the
text.  

Let $N(\gamma)$ be the number density per unit Lorentz factor of relativistic
electrons defined as non-zero between some minimum and maximum Lorentz factor
$\gamma_{\rm{min}}$ and $\gamma_{\rm{max}}$ in a magnetic field of flux
density B. Further, define the non-relativistic gyro-frequency as $\Omega_0$,
and the classical electron radius as $r_e$. Written in S.I. units, $\Omega_0 =
\frac{q_e B}{m_e}$, and in c.g.s. units, $\Omega_0 = q_e B/m_e c$ where $q_e$
and $m_e$ are the electron charge and mass respectively. Let $y = (4 \pi / 3)
\Omega_0^{-1} \gamma^{-2}$. The angle averaged synchrotron emissivity (valid
in both S.I. and c.g.s. units) is 

\begin{eqnarray} \label{eqn:synch_emissivity}
\langle j_{\nu} \rangle &=& \frac{1}{4 \pi}\int_{4 \pi} j_{\nu} d\Omega \\
 &=& \left( \frac{m_e r_e c}{4 \pi^{1/2}}  \right) \nu^{1/2} \Omega_0^{1/2} 
\int_{y_1}^{y_2} y^{-3/2} N(\gamma(y)) \bar{F}(y) dy  \label{eqn:j_nu_av} \\
\end{eqnarray}
where
\begin{eqnarray}
\bar{F}(y) &=& \int_0^{\pi/2}F \left( \frac{y}{\sin \psi} 
\right) \sin^2 \psi d\psi \\
&=& y\int_y^{\infty} \left(1 - \frac{y^2}{t^2} \right)^{1/2} K_{5/3}(t)dt
\end{eqnarray}
and the synchrotron function $F(x) = x \int_x^{\infty} K_{5/3}(z)dz$, where
$K_{5/3}(z)$ is the modified Bessel function of order 5/3. The integration
limits on y are given by 

\begin{eqnarray} 
y_1 &=& \frac{4 \pi}{3} \frac{\nu}{\Omega_0} \gamma_{\rm{max}}^{-2} = \nu/\nu_2 \\
y_2 &=& \frac{4 \pi}{3} \frac{\nu}{\Omega_0} \gamma_{\rm{min}}^{-2} = \nu/\nu_1
\end{eqnarray}
where $\nu_1$ and $\nu_2$ are characteristic frequencies corresponding to
$\gamma_{\rm{min}}$ and $\gamma_{\rm{max}}$, viz. $\frac{3}{4 \pi} \Omega_0
\gamma_{\rm{min, max}}^2$.  

Using the volume averaged shock distribution function $\bar{N}(\gamma(y))$
defined by equations (\ref{eqn:N_gamma}) and (\ref{eqn:g_gamma}) the angle
averaged emission spectrum is

\begin{equation} \label{eqn:j_nu_syn}
\langle F_{\nu} \rangle =  \frac{\delta^{(a+3)/2} (1+z)^{(3-a)/2} V}{D_L^2}
A(a) \nu^{-a/2} \nu_b^{1/2} A_{\rm{syn}} \int_{\nu/\nu_2}^{\nu/\nu_1}
y^{\frac{a-2}{2}} \bar{F}(y) \; g \! \left( y, \frac{\nu}{\nu_b}  \right) dy  
\end{equation}
where
\begin{eqnarray}
g \left( y, \frac{\nu}{\nu_b}  \right) &=& \left \{ \begin{array}{ll} 1 -
  \left[ 1 - \left( \frac{\nu}{y \nu_b}  \right)^{1/2} \right]^{a-1}  &  y >
  \frac{\nu}{\nu_b} \\ 
1 &  y < \frac{\nu}{\nu_b} \\
\end{array}
\right. \\
\nu_{1, b, 2} &=& \frac{\delta}{(1+z)}\frac{3}{4 \pi}  \Omega_0
\gamma_{\rm{min, b, max}}^2  \\ 
A_{\rm{syn}} &=& K_e \Omega_0^{\frac{a+1}{2}} \\
A(a) &=& \frac{r_e m_e c \; 3^{a/2}}{(a-1) \; 2^{a+2} \, \pi^{\frac{a+1}{2}}}
\end{eqnarray}
The shape of the model synchrotron spectrum is determined by the parameters
$\nu_1$, $\nu_b$, $\nu_2$ and a. The amplitude is governed by the parameter
$A_{\rm{syn}}$. In equation (\ref{eqn:j_nu_syn}) we have assumed the emission
is produced by plasma flowing at a relativistic speed $\beta c$ at an angle
$\theta$ to the line of sight with corresponding Lorentz factor $\Gamma =
\left( 1 - \beta^2  \right)^{-1/2}$ and Doppler factor $\delta = \left[ \Gamma
  (1 - \beta \cos \theta) \right]^{-1}$ through a stationary volume or
pattern, so that $F_{\nu} \propto \delta^{(a+3)/2}$ as appropriate for
extragalactic jets \citep{lind85}. If the volume in which the flux is produced
is moving relativistically, an extra factor of $\delta$ enters, so that the
leading factor of $\delta^{(a+3)/2}$ in equation (\ref{eqn:j_nu_syn}) becomes
$\delta^{(a+5)/2}$.  

To calculate the model synchrotron spectrum, we first specify a source volume
V, Doppler factor $\delta$, redshift z and corresponding luminosity distance
$D_L$. We then specify lab frame values for the critical frequencies $\nu_1$,
$\nu_b$ and $\nu_2$ and the observation frequency $\nu$. Finally, we specify
the synchrotron amplitude $A_{\rm{syn}}$, and calculate the spectrum
numerically using equation (\ref{eqn:j_nu_syn}).  

\subsection{Synchrotron Self Compton Flux Density}

Let $\epsilon_s$ be the inverse Compton scattered photon energy, $\epsilon_i$
the soft photon energy,  $n(\epsilon_i)$ the number density of soft photons
per unit energy, $N(\gamma)$ the number density of relativistic electrons per
unit Lorentz factor and $\sigma_T$ the Thompson cross section. The inverse
Compton emissivity (in the Thompson limit) from an isotropic distribution of
relativistic electrons in an isotropic soft photon field is given by  

\begin{eqnarray}
j_{\epsilon_s} &=& \frac{3 c \sigma_T}{16 \pi} \epsilon_s \int_0^{\infty}
\frac{n\left( \epsilon_i \right)}{\epsilon_i} \left[
  \int_{\gamma_{\rm{min}}}^{\gamma_{\rm{max}}} \frac{N_e\left( \gamma
    \right)}{\gamma^2} F_C(q) d \gamma \right] d \epsilon_i \\ 
\vspace{0.1in}
\mbox{where} \hspace{0.2in} q &=& \frac{\epsilon_s}{4 \epsilon_i \gamma^2} \\
\mbox{and}  \hspace{0.2in} F_C(q) &=& 2 q \, \rm{ln} (q) + q + 1 - 2 q^2 \nonumber
\end{eqnarray}
\citep[eg.][]{blumenthal70}. This expression is valid for any particle
distribution and any photon distribution, provided they are both
isotropic. Changing the integration variable from $\gamma$ to q, we obtain 

\begin{eqnarray}
j_{\epsilon_s} &=& \frac{3 c \sigma_T}{2^6 \pi} \epsilon_s^{3/2}
\int_0^{\infty} \frac{n\left(  \epsilon_i \right)}{\epsilon_i^{3/2}} \left[
  \int_{q2}^{q1} \frac{N(\gamma)}{\gamma^2} q^{-3/2} F_C(q) dq \right]
d\epsilon_i  
\end{eqnarray}
where
\begin{equation}
q_{1, 2} = \frac{\epsilon_s}{4 \epsilon_i \gamma^2_{\rm{min, max}}}
\end{equation}

Using the broken power law distribution described by equations
(\ref{eqn:N_gamma}) and (\ref{eqn:g_gamma}) and expressing it in terms of the
variable q, we find  

\begin{equation} \label{eqn:F_nu_ssc_II}
F^{\rm{ssc}}_{\nu} = \left(  \frac{9 \cdot 2^{(a-3)} R \; \sigma_T}{a-1}
\right)  \nu^{-a/2} A_{\rm{ssc}} \int_{\nu_1}^{\nu_2} \nu_i^{(a-2)/2}
F_{\nu_i}^{\rm{syn}} \left[  \int_{q_2}^{q_1} q^{a/2} F_C(q) g \left(
  \frac{q_b}{q} \right) dq \right] d\nu_i 
\end{equation}
where
\begin{eqnarray}
A_{\rm{ssc}} &=& K_e \gamma_b \nonumber \\
q_b &=& \frac{\epsilon_s}{4 \epsilon_i \gamma^2_b} \\
g \left( \frac{q_b}{q} \right)  &=& \left \{ \begin{array}{ll}  1 - \left(1 -
  \left( \frac{q_b}{q} \right)^{1/2} \right)^{a-1}    &  q > q_b \\ 
1 &  q < q_b \\
\end{array}
\right.
\end{eqnarray}
and $F_{\nu_i}^{\rm{syn}}$ is calculated using equation
(\ref{eqn:j_nu_syn}). In equation (\ref{eqn:F_nu_ssc_II}) we have assumed 
\begin{eqnarray} \label{eqn:n_epsilon}
\int n(\epsilon^{\prime}_i) dV^{\prime} &=& \frac{3 \pi}{h^2 \nu_i^{\prime} c}
R j^{\prime \, \rm{syn}}_{\nu_i^{\prime}} V^{\prime} \\
&=& \frac{3 \pi R}{h^2 \nu_i c} \frac{F^{\rm{syn}}_{\nu_i} D_L^2}{\delta (1+z)} \label{eqn:n_eps_int}
\end{eqnarray}
as appropriate for a spherical region of homogenous plasma. In the above
expression $\epsilon_{i}^{\prime}$ and $\nu_i^{\prime}$ are the incident
photon energy and frequency in the rest frame of the plasma, and
$j_{\nu_i}^{\rm{syn}}$ is the synchrotron emissivity at frequency
$\nu_i^{\prime}$ in the rest frame of the plasma.  

We calculate the synchrotron self Compton spectrum by specifying the value of
$A_{\rm{ssc}}$ along with the best fit values for $A_{\rm{syn}}$, $\nu_1$,
$\nu_b$ and $\nu_2$ determined from fitting the synchrotron spectrum. The
spectrum is calculated using equations (\ref{eqn:F_nu_ssc_II}). Together, the
five parameters $\nu_1$, $\nu_b$, $\nu_2$, $A_{\rm{syn}}$ and $A_{\rm{ssc}}$
allow the parameters $K_e$, B, $\gamma_{\rm{min}}$, $\gamma_b$ and
$\gamma_{\rm{max}}$ to be determined.


\begin{thebibliography}{}

\bibitem[Aloy et al.(1999)]{aloy99} Aloy, M.~A., 
Ib{\'a}{\~n}ez, J.~M.~\^{}., Mart{\'{\i}}, J.~M.~\^{}., G{\'o}mez, J.-L., 
M\"{u}ller, E.\ 1999, \apjl, 523, L125 

\bibitem[Amato 
\& Arons(2006)]{amato06} Amato, E., \& Arons, J.\ 2006, \apj, 653, 325 

\bibitem[Arshakian 
\& Longair(2000)]{arshakian00} Arshakian, T.~G., \& Longair, M.~S.\ 2000, \mnras, 311, 846 

\bibitem[Ba{\l}uci{\'n}ska-Church et al.(2005)]{balucinska-church05} 
Ba{\l}uci{\'n}ska-Church, M., Ostrowski, M., Stawarz, {\l}., 
\& Church, M.~J.\ 2005, \mnras, 357, L6 

\bibitem[Belsole et al.(2006)]{belsole06} Belsole, E., Worrall, 
D.~M., \& Hardcastle, M.~J.\ 2006, \mnras, 366, 339 

\bibitem[Blumenthal 
\& Gould(1970)]{blumenthal70} Blumenthal, G.~R., \& Gould, R.~J.\ 1970, Reviews of Modern Physics, 42, 237 

\bibitem[Blundell et al.(2006)]{blundell06} Blundell, K.~M., 
Fabian, A.~C., Crawford, C.~S., Erlund, M.~C., 
\& Celotti, A.\ 2006, \apjl, 644, L13 

\bibitem[Bridle et al.(1994)]{bridle94} Bridle, A.~H., Hough, 
D.~H., Lonsdale, C.~J., Burns, J.~O., \& Laing, R.~A.\ 1994, \aj, 108, 766 

\bibitem[Brunetti et al.(2003)]{brunetti03} Brunetti, G., Mack, 
K.-H., Prieto, M.~A., \& Varano, S.\ 2003, \mnras, 345, L40 

\bibitem[Campbell-Wilson \& Hunstead(1994)]{campbell-wilson94} 
Campbell-Wilson, D., \& Hunstead, R.~W.\ 1994, Proceedings of the 
Astronomical Society of Australia, 11, 33 

\bibitem[Carilli et al.(1988)]{carilli88} Carilli, C.~L., Dreher, 
J.~W., Perley, R.~A., Leahy, P., \& Muxlow, T.\ 1988, \baas, 20, 734 

\bibitem[Carilli et al.(1991)]{carilli91} Carilli, C.~L., Perley, 
R.~A., Dreher, J.~W., \& Leahy, J.~P.\ 1991, \apj, 383, 554 

\bibitem[Carilli et al.(1999)]{carilli99} Carilli, C.~L., Kurk, 
J.~D., van der Werf, P.~P., Perley, R.~A., 
\& Miley, G.~K.\ 1999, \aj, 118, 2581 

\bibitem[Cheung et al.(2005)]{cheung05} Cheung, C.~C., Wardle, 
J.~F.~C., \& Chen, T.\ 2005, \apj, 628, 104 

\bibitem[Dennett-Thorpe et al.(1997)]{dennett-thorpe97} Dennett-Thorpe, 
J., Bridle, A.~H., Scheuer, P.~A.~G., Laing, R.~A., \& Leahy, J.~P.\ 1997, 
\mnras, 289, 753

\bibitem[de Young(2002)]{deyoung02} de Young, D.~S.\ 2002, The 
physics of extragalactic radio sources, by David S.~De Young.~Chicago, 
Ill.~: University of Chicago Press, 2002.,  

\bibitem[Double et al.(2004)]{double04} Double, G.~P., Baring, 
M.~G., Jones, F.~C., \& Ellison, D.~C.\ 2004, \apj, 600, 485 

\bibitem[Ekers(1969)]{ekers69} Ekers, J.~A.\ 1969, Australian 
Journal of Physics Astrophysical Supplement, 7, 3 

\bibitem[Erlund et al.(2008)]{erlund08} Erlund, M.~C., Fabian, 
A.~C., \& Blundell, K.~M.\ 2008, \mnras, 386, 1774 

\bibitem[Francis et al.(1991)]{francis91} Francis, P.~J., Hewett, 
P.~C., Foltz, C.~B., Chaffee, F.~H., Weymann, R.~J., \& Morris, S.~L.\ 
1991, \apj, 373, 465

\bibitem[Gambill et 
al.(2003)]{gambill03} Gambill, J.~K., Sambruna, R.~M., Chartas, G., Cheung, C.~C., Maraschi, L., Tavecchio, F., Urry, C.~M., \& Pesce, J.~E.\ 2003, \aap, 401, 505 

\bibitem[Gelbord et al.(2005)]{gelbord05} Gelbord, J.~M., et al.\ 
2005, \apjl, 632, L75 (G05)

\bibitem[Georganopoulos \& Kazanas(2003)]{georganopoulos03} 
Georganopoulos, M., \& Kazanas, D.\ 2003, \apjl, 589, L5

\bibitem[Gopal-Krishna et al.(2004)]{gopal-krishna04} Gopal-Krishna, 
Biermann, P.~L., \& Wiita, P.~J.\ 2004, \apjl, 603, L9 

\bibitem[Hardcastle et al.(1998)]{hardcastle98} Hardcastle, M.~J., 
Alexander, P., Pooley, G.~G., \& Riley, J.~M.\ 1998, \mnras, 296, 445 

\bibitem[Hardcastle et al.(2001a)]{hardcastle01a} Hardcastle, M.~J., 
Birkinshaw, M., \& Worrall, D.~M.\ 2001, \mnras, 323, L17 

\bibitem[Hardcastle(2001b)]{hardcastle01b} Hardcastle, M.~J.\ 2001b, \aap, 373, 881 

\bibitem[Hardcastle et al.(2002)]{hardcastle02} Hardcastle, M.~J., 
Birkinshaw, M., Cameron, R.~A., Harris, D.~E., Looney, L.~W., \& Worrall, 
D.~M.\ 2002, \apj, 581, 948

\bibitem[Hardcastle(2003)]{hardcastle03} Hardcastle, M.~J.\ 2003, 
New Astronomy Review, 47, 649

\bibitem[Hardcastle et al.(2004)]{hardcastle04} Hardcastle, M.~J., 
Harris, D.~E., Worrall, D.~M., \& Birkinshaw, M.\ 2004, \apj, 612, 729 


\bibitem[Harris et al.(2000)]{harris00} Harris, D.~E., et al.\ 
2000, \apjl, 530, L81

\bibitem[Heavens \& Meisenheimer(1987)]{heavens87} Heavens, 
A.~F., \& Meisenheimer, K.\ 1987, \mnras, 225, 335

\bibitem[Hoshino et al.(1992)]{hoshino92} Hoshino, M., Arons, J., 
Gallant, Y.~A., \& Langdon, A.~B.\ 1992, \apj, 390, 454 

\bibitem[Ishwara-Chandra 
\& Saikia(2000)]{ishwara-chandra00} Ishwara-Chandra, C.~H., \& Saikia, D.~J.\ 2000, \mnras, 317, 658 

\bibitem[Jeyakumar \& Saikia(2000)]{jeyakumar00} Jeyakumar, S., \& 
Saikia, D.~J.\ 2000, \mnras, 311, 397

\bibitem[Jones 
\& Odell(1977)]{jones77} Jones, T.~W., \& Odell, S.~L.\ 1977, \aap, 61, 291 

\bibitem[Kataoka 
\& Stawarz(2005)]{kataoka05} Kataoka, J., \& Stawarz, {\L}.\ 2005, \apj, 622, 797 

\bibitem[Komissarov 
\& Falle(1996)]{komissarov96} Komissarov, S.~S., \& Falle, S.~A.~E.~G.\ 1996, Energy Transport in Radio Galaxies and Quasars, 100, 327 

\bibitem[K\"onigl(1980)]{konigl80} K\"onigl, A.\ 1980, Phys. Fluids, 23, 1083 

\bibitem[Large et al.(1981)]{large81} Large, M.~I., Mills, 
B.~Y., Little, A.~G., Crawford, D.~F., 
\& Sutton, J.~M.\ 1981, \mnras, 194, 693 

\bibitem[Lazio et al.(2006)]{lazio06} Lazio, T.~J.~W., Cohen, 
A.~S., Kassim, N.~E., Perley, R.~A., Erickson, W.~C., Carilli, C.~L., \& 
Crane, P.~C.\ 2006, \apjl, 642, L33 

\bibitem[Leahy et al.(1989)]{leahy89} Leahy, J.~P., Muxlow, 
T.~W.~B., \& Stephens, P.~W.\ 1989, \mnras, 239, 401 

\bibitem[Lind 
\& Blandford(1985)]{lind85} Lind, K.~R., \& Blandford, R.~D.\ 1985, \apj, 295, 358 

\bibitem[Linfield et al.(1989)]{linfield89} Linfield, R.~P., et 
al.\ 1989, \apj, 336, 1105 

\bibitem[Lovell(1997)]{lovell97} Lovell, J.~E.~J.\ 1997, 
Ph.D. thesis, Univ. Tasmania

\bibitem[Marscher(1988)]{marscher88} Marscher, A.~P.\ 1988, \apj, 
334, 552 

\bibitem[Marshall et al.(2005)]{marshall05} Marshall, H.~L., et 
al.\ 2005, \apjs, 156, 13 

\bibitem[Meisenheimer et 
al.(1997)]{meisenheimer97} Meisenheimer, K., Yates, M.~G., \& Roeser, H.-J.\ 1997, \aap, 325, 57 

\bibitem[Meisenheimer(2003)]{meisenheimer03} Meisenheimer, K.\ 2003, 
New Astronomy Review, 47, 495 

\bibitem[Norman(1996)]{norman96} Norman, M.~L.\ 1996, Energy 
Transport in Radio Galaxies and Quasars, 100, 319

\bibitem[Preston et al.(1989)]{preston89} Preston, R.~A., et al.\ 
1989, \aj, 98, 1

\bibitem[Prieto et al.(2002)]{prieto02} Prieto, M.~A., Brunetti, 
G., \& Mack, K.-H.\ 2002, Science, 298, 193 

\bibitem[Rayner et al.(2000)]{rayner00} Rayner, D.~P., Norris, 
R.~P., \& Sault, R.~J.\ 2000, \mnras, 319, 484 

\bibitem[Roeser 
\& Meisenheimer(1987)]{roser87} Roeser, H.-J., \& Meisenheimer, K.\ 1987, \apj, 314, 70 

\bibitem[Scheuer(1995)]{scheuer95} Scheuer, P.~A.~G.\ 1995, 
\mnras, 277, 331

\bibitem[Schwartz et al.(2006)]{schwartz06} Schwartz, D.~A., et 
al.\ 2006, \apj, 640, 592 

\bibitem[Shepherd(1997)]{shepherd97} Shepherd, M.~C.\ 1997, 
Astronomical Data Analysis Software and Systems VI, 125, 77

\bibitem[Spitkovsky(2008)]{spitkovsky08} Spitkovsky, A.\ 2008, 
\apjl, 673, L39 

\bibitem[Stawarz et al.(2007)]{stawarz07} Stawarz, {\L}., Cheung, 
C.~C., Harris, D.~E., \& Ostrowski, M.\ 2007, \apj, 662, 213 

\bibitem[Tavecchio et al.(2000)]{tavecchio00} Tavecchio, F., 
Maraschi, L., Sambruna, R.~M., \& Urry, C.~M.\ 2000, \apjl, 544, L23 

\bibitem[Tavecchio et al.(2005)]{tavecchio05} Tavecchio, F., 
Cerutti, R., Maraschi, L., Sambruna, R.~M., Gambill, J.~K., Cheung, C.~C., 
\& Urry, C.~M.\ 2005, \apj, 630, 721 

\bibitem[Tsang 
\& Kirk(2007)]{tsang07} Tsang, O., \& Kirk, J.~G.\ 2007, \aap, 463, 145 

\bibitem[Tregillis et al.(2001)]{tregillis01} Tregillis, I.~L., 
Jones, T.~W., \& Ryu, D.\ 2001, \apj, 557, 475

\bibitem[Uchiyama et al.(2005)]{uchiyama05} Uchiyama, Y., Urry, 
C.~M., Van Duyne, J., Cheung, C.~C., Sambruna, R.~M., Takahashi, T., 
Tavecchio, F., \& Maraschi, L.\ 2005, \apjl, 631, L113 

\bibitem[Urry \& Padovani(1995)]{urry95} Urry, C.~M., \& 
Padovani, P.\ 1995, \pasp, 107, 803

\bibitem[Wardle(1977)]{wardle77} Wardle, J.~F.~C.\ 1977, \nat, 
269, 563 

\bibitem[Wardle et al.(1998)]{wardle98} Wardle, J.~F.~C., Homan, 
D.~C., Ojha, R., \& Roberts, D.~H.\ 1998, \nat, 395, 457 

\bibitem[Wills(1975)]{wills75} Wills, B.~J.\ 1975, Australian 
Journal of Physics Astrophysical Supplement, 38, 1 

\bibitem[Wilson et al.(2006)]{wilson06} Wilson, A.~S., Smith, 
D.~A., \& Young, A.~J.\ 2006, \apjl, 644, L9

\bibitem[Worrall 
\& Birkinshaw(2006)]{worrall06} Worrall, D.~M., \& Birkinshaw, M.\ 2006, Physics of Active Galactic Nuclei at all Scales, Lecture Notes in Physics, 693, 39 


\end{thebibliography}
\end{document}